\documentclass[twocolumn,aps,epsfig,nofootinbib,preprintnumbers]{revtex4}

\usepackage{graphicx}
\usepackage{epstopdf}
\usepackage{latexsym}
\usepackage{amssymb}
\usepackage{amsmath}
\usepackage{color}
\usepackage{mathrsfs}

\usepackage[center]{subfigure}

\begin{document}

 \newcommand{\bq}{\begin{equation}}
 \newcommand{\eq}{\end{equation}}
 \newcommand{\bqn}{\begin{eqnarray}}
 \newcommand{\eqn}{\end{eqnarray}}
 \newcommand{\nb}{\nonumber}
 \newcommand{\lb}{\label}
\newcommand{\PRL}{Phys. Rev. Lett.}
\newcommand{\PL}{Phys. Lett.}
\newcommand{\PR}{Phys. Rev.}
\newcommand{\CQG}{Class. Quantum Grav.}

%
\title{Phenomenological implications of modified loop cosmologies: an overview} 

\author{Bao-Fei Li $^{1, 2}$\footnote{E-mail address: baofeili1@lsu.edu}}
\author{Parampreet Singh$^1$\footnote{ E-mail address: psingh@lsu.edu}}
\author{Anzhong Wang$^{3}$\footnote{E-mail address: Anzhong$\_$Wang@baylor.edu}}
\affiliation{$^1$ Department of Physics and Astronomy,  
Louisiana State University, Baton Rouge, LA 70803, USA\\
$^{2}$ Institute for Theoretical Physics $\&$ Cosmology, Zhejiang University of Technology, Hangzhou, 310032, China\\
$^3$ GCAP-CASPER, Department of Physics, Baylor University, Waco, TX, 76798-7316, USA
}

\date{\today}

\begin{abstract}

 In this  paper,  we first provide a brief review of the effective dynamics of two recently well-studied  models of modified loop 
quantum cosmologies (mLQCs), which arise from different regularizations of the Hamiltonian constraint and show the robustness 
of a generic resolution of the big bang singularity, replaced by a quantum bounce due to non-perturbative Planck scale effects. 
As in loop quantum cosmology (LQC), in  these modified models the slow-roll inflation happens generically.  We consider 
the cosmological perturbations  following the dressed and hybrid approaches and  clarify some subtle issues regarding the 
ambiguity of the extension of the effective potential of the scalar perturbations across the quantum bounce,  and the choice of initial 
conditions. Both of the  modified regularizations yield primordial power spectra that are consistent with current observations 
for the Starobinsky potential within the framework of either the dressed or the hybrid approach.  But differences in primordial 
power spectra are  also identified  among the mLQCs and LQC.  In  particular,  for mLQC-I, striking differences arise between 
the dressed and hybrid approaches in the infrared and oscillatory regimes.  While the differences between the two modified models
can be attributed to  differences in the Planck scale physics, the permissible choices of the initial conditions and the differences
between the two perturbation approaches  have been reported for the first time. All these differences,  due to either the different  
regularizations or the different perturbation approaches   {,}  in principle can be observed  in terms of non-Gaussianities.

\end{abstract}
%
%
\maketitle

\section{Introduction}
\label{Intro}
\renewcommand{\theequation}{1.\arabic{equation}}\setcounter{equation}{0}

Despite offering a solution to several fundamental and conceptual problems of the standard big bang cosmology, 
 including the flatness, horizon, and exotic-relic problems, the cosmic inflation in the early universe
 also provides a  mechanism to produce  density perturbations and primordial gravitational waves (PGWs) \cite{Guth,inflation,InfReviews}. The latter arise from quantum fluctuations of  spacetimes and 
produce not only a temperature anisotropy, but also polarizations  in  the cosmic microwave background (CMB), a smoking gun of PGWs.
However, the inflationary paradigm is incomplete without the knowledge of key elements from quantum gravity. First, it is well-known that the cosmic inflation is sensitive to the ultraviolet (UV) physics, and its successes are tightly contingent on the understanding of this UV physics \cite{trans-planck,KW14,stringC}. 
In particular,  if the inflationary phase lasts somewhat longer than the minimal period required to solve the above mentioned problems,  the length scales we observe today will originate from modes that are smaller
 than the Planck length during inflation \cite{trans-planck}. Then, the treatment of the underlying quantum field theory on a classical spacetime becomes questionable, as now the quantum geometric effects are expected to be large, and  the space and time cannot be treated  classically  any more.
 This is often referred to as   {\em the trans-Planckian problem} of cosmological fluctuations \footnote{It has been conjectured using models in string theory that the 
 trans-Planckian problem might never arise \cite{BV20}, which results on severe constraints on various cosmological models (See \cite{BBLV20,Brand20} for more details).}.
  
The second problem of the inflationary paradigm is more severe. It is well known that inflationary spacetimes are past-incomplete because of the big bang singularity \cite{singularity}, with which it is not clear how to impose the  initial conditions. This problem gets aggravated for low energy inflation in spatially-closed models which are slightly favored by current observations where the universe encounters a big crunch singularity and lasts only for a few Planck seconds \cite{linde}.  

Another problematic feature of inflation is that  one often   ignores the pre-inflationary dynamics and 
sets  the Bunch-Davies (BD) vacuum at a very early time. But, 
 it is not clear how such a vacuum state can be realized dynamically  in the framework of quantum cosmology \cite{stringC}, considering the fact that a pre-inflationary phase  always exists between the Plank
and inflation scales, which are about $10^3 \sim 10^{12}$ orders of magnitude difference \cite{inflation,stringC}.  
While these problems of inflationary paradigm demand a completion from quantum theory of spacetimes, they also open an avenue to overcome one of the  main obstacles    in the development of quantum gravity,
 which concerns with   the  lack of  experimental evidences. Thus, understanding inflation in the framework of quantum gravity could offer valuable guidances to the construction of the underlying theory \cite{QGs,string,LQG}. 
 
 In particular, when applying the techniques of loop quantum gravity (LQG) to homogeneous and isotropic universe, namely   loop quantum cosmology (LQC) \cite{LQC}, it was shown that,
 purely due to  quantum geometric effects,  the big bang singularity is generically resolved  and replaced by a quantum bounce  at which the  spacetime curvature becomes Planckian \cite{aps1,aps2,aps3,acs2010}. The robustness of the singularity resolution has been shown for a variety of isotropic and anisotropic spacetimes \cite{giesel-li-singh-inflation}.   Interestingly, there exists a reliable effective spacetime description,
  which has been used to confirm a generic resolution of all strong curvature singularities \cite{non-singular}.  Various phenomenological implications have been studied using this effective spacetime description,
  whose validity has been verified for isotropic and anisotropic spacetimes \cite{numlsu-1,numlsu-2,numlsu-3,numlsu-4,numlsu-5}. For low energy inflation models with a positive spatial curvature,  
  the singularity resolution  and a successful onset of inflation for  classically inadmissible initial conditions have been demonstrated \cite{dupuy-singh,gordon-li-singh,warm-inflation}.

  In the last couple of years,  several approaches have been proposed, in order to address the impacts of the quantum geometry on the primordial power spectra. These include
 the approaches of the deformed algebra  \cite{bhks2008,cbgv2012,cmbg2012},  
  dressed metric  \cite{aan2012,aan2013,aan2013b},
  and  hybrid  \cite{mm2012,mm2013,gmmo2014,gbm2015,mo2016}
  (For a recent discussion about similar ideas in anisotropic Bianchi I LQC spacetimes see Refs. \cite{GS12,GS13,b1-lett,b1-long} and references therein.).  In particular,  the last two approaches have been widely studied and  found that they are all consistent with current cosmological observations \cite{agullo_detailed_2015,d1,d1b,bonga_inflation_2016,bonga_phenomenological_2016,ZhuA,ZhuB,ZWKCS18,WZW18,abs2018,bo2016,gbmo2016,nbm2018,nbm2018a}.
  In addition, within the framework of  the dressed metric approach recently  it has been also shown that  some anomalies from the CMB data \cite{Planck2018b,Planck2019,SCHS16} can be
  reconciled  purely due to the quantum geometric effects \cite{AGJS20,AKS20}.

In addition to the standard LQC, in which the Lorentzian term  of the classical  Hamiltonian constraint is  first expressed in terms of the Euclidean term in the spatially flat
 Friedmann-Lema\^itre-Robertson-Walker (FLRW) universe, and then only the quantization of the Euclidean term is considered,  the robustness of the singularity resolution with respect to different  quantizations of the classical Hamiltonian constraint in the symmetry reduced spacetimes have been extensively studied. Two notable examples are the so-called  modified LQC-I (mLQC-I)  and modified LQC-II (mLQC-II) models,  which were first proposed by Yang, Ding and Ma more than a decade ago \cite{YDM09}.
In a recent study, Dapor and Liegener (DL) \cite{DL17,DL18} obtained the expectation values of the Hamiltonian operator in LQG using complexifier coherent states \cite{states}, adapted to the spatially flat  FLRW
 universe.   Using the non-graph changing regularization of the Hamiltonian advocated by Thiemann \cite{thiemann}, DL obtained an effective Hamiltonian constraint, which,  to the leading order,  agrees with the mLQC-I model first obtained in  \cite{YDM09}. Sometimes, this model has also been  referred to as the DL model or Thiemann regularized LQC.  Strictly speaking, 
 when constructing loops  in \cite{DL17} DL treated the edge length $\mu$  as a free parameter, but in \cite{YDM09} it was considered as a specific triad dependent function, the so-called $\bar \mu$ scheme \cite{aps3},  which  is known to be the only possible choice in LQC, and results in physics that is independent from underlying fiducial structures used during quantization, and meanwhile yields a consistent infrared behavior for all matter obeying the weak energy condition \cite{cs08}. Lately,  the studies of \cite{DL17}  have been extended to the $\bar \mu$ scheme \cite{adlp}.

 In the two modified LQC models, mLQC-I  and mLQC-II,  since different regularizations of the Lorentzian term  were used,
   the resulting  equations become the fourth-order and non-singular quantum difference equations, instead of 
 the second-order  difference ones obtained in LQC. In these two models the big bang singularity is also generically resolved and replaced by a quantum bounce. In addition,   the inflationary phase can naturally take place with a very high probability  \cite{lsw2018,lsw2018b,lsw2019,SS19a,SS19b}. In addition, the dynamics in LQC and mLQC-II is qualitatively similar in the whole evolution of the universe, while the one in mLQC-I becomes significantly different from LQC (as well as mLQC-II)  in the contracting phase, in which an emergent quasi de Sitter space is present. This implies that  the contracting phase in mLQC-I is purely a quantum regime without any classical limit \footnote{A similar contracting branch is found in certain anisotropic models in the standard regularization of LQC (see for eg. \cite{dadhich-joe-singh-kantowski-sachs}).}.
 
 An important question now is what are the effects of these models and approaches on the CMB observations.  The answer to this question requires the knowledge of how the quantum fluctuations 
propagate on a quantum spacetime in LQC and modified loop cosmological models. 
In particular,  in the framework of  the dressed metric approach  the power spectra of the cosmological perturbations for
 both mLQC-I and mLQC-II models were investigated \cite{lsw2020}. 
 In  the same framework but restricted only to the mLQC-I  model, the power spectra of the cosmological perturbations  were   studied in \cite{IA19}. 
 More recently, the hybrid approach was applied to  mLQC-I \cite{qm2019,gm20,gqm2020}, for which the time-dependent  mass of the perturbations   was studied  in detail  \cite{qmp2020}.
 The primordial scalar power spectra obtained in the three models, LQC, mLQC-I and mLQC-II, were also investigated in the hybrid approach  \cite{LOSW20}, and  found that the relative differences in the amplitudes of the power spectra among the three models could be as large as $2\%$ in the UV regime of the spectra, which is relevant to the current observations.   Interestingly, in the above work, differences in primordial power spectra were found between the hybrid and dressed metric approaches in the infra-red and oscillatory regimes in mLQC-I.

In this brief review, we shall focus mainly on the states that are sharply peaked along the classical trajectories, so that
the description of the ``effective" dynamics of the universe becomes available \cite{LQC}, and the questions raised recently in  \cite{KKL20} are avoided. This includes the studies of
the ``effective" dynamics of the homogeneous and isotropic mLQC-I  and mLQC-II models, and their cosmological  perturbations in the framework of the
dressed metric and hybrid approaches. We shall first clarify the issue regarding the ambiguities  in the extension of the effective potential for the scalar perturbations across the quantum bounce, 
and then pay particular attention to  the differences among the three models, LQC, mLQC-I  and mLQC-II,
and  possible observational signals. It is important to note that initial conditions are another subtle and important issue not only in LQC but also in mLQCs. This includes two parts: (i) when to impose the initial  conditions, and (ii)
which kind of initial conditions one can impose {\em consistently}. To clarify this issue, we discuss  it at length by showing the  (generalized) comoving Hubble radius in each model
 and in each of  the dressed and  hybrid approaches. From this analysis, one can see clearly what initial conditions can and cannot be imposed at a chosen initial time.  

The outline of this brief overview is as follows.  In Sec. \ref{SecII} we consider the effective dynamics of mLQC-I and mLQC-II, and discuss some universal features of their dynamics such as the resolution of big bang singularity. 
In addition, in this section  we also show that for states such that the evolution of the homogeneous universe was dominated initially at the bounce by the kinetic energy of the inflaton, that is, 
$\dot\phi_B^2 \gg V(\phi_B)$,
 the post-bounce evolution between the bounce and the reheating can be always divided universally into three different phases: {\em the bouncing, transition, 
and slow-roll inflation}  [cf. Fig. \ref{fig1}]. During 
each of these phases the expansion factor $a(t)$ and the scalar field $\phi(t)$ can be given analytically. In particular, they are given by Eqs.(\ref{3.21})-(\ref{3.22}) and (\ref{3.23})-(\ref{3.24}) during the bouncing phase
for mLQC-I and mLQC-II, respectively. In this same section, the probabilities of the slow-roll inflation is considered, and shown that it occurs generically. This particular consideration is restricted to the quadratic potential, but is expected to be  also true for other cases.  

In Sec. III, the cosmological perturbations of mLQC-I and mLQC-II are studied. We discuss initial
conditions and the subtle issue of the ambiguity in the choice of the variables $\pi_a^{-2}$ and $\pi_a^{-1}$ (present in the effective potential), which correspond to the quadratic and linear
inverse of the momentum conjugate to the
scale factor. In  addition, to understand the issue of initial conditions
 properly, we first introduce the comoving Hubble radius $\lambda_H^2$ and then state clearly how this is resolved in GR [cf. Fig. \ref{fig2}],
and which are the relevant questions in mLQC-I [cf. Fig. \ref{fig5}] and   mLQC-II [cf. Fig. \ref{fig6}]. From these figures it is clear that  the BD vacuum  cannot be
consistently imposed at the bounce  \footnote{It should be noted that anisotropies rise during
the contracting phase and generically dominate the earliest stages of the post-bounce of the homogeneous universe \cite{GS12,GS13,b1-lett,b1-long}. So, cautions must be taken,
when imposing initial conditions at the bounce.}, as now some modes are inside the (comoving) Hubble radius while others not. However, the fourth-order adiabatic vacuum  may be  imposed at this moment for both of these
two modified LQC models, as that adopted in  LQC \cite{aan2013b}. In addition,  in mLQC-I  the de Sitter state given by Eq.(\ref{5.2b}) \footnote{To be distinguished from the BD vacuum
described by Eq.(\ref{5.3}) we refer to  the state described by Eq.(\ref{5.2b}) as the de Sitter state. The difference between them is due to the term $i/(k\eta)$, which is not negligible in the deep contracting phase of the de Sitter background, as now $|k\eta|$ could be very small. For more details, see the discussions given in Sec. III.A, especially the paragraph after Eq.(\ref{5.6}).}  can be imposed in the contracting phase as long as $t_0$ is sufficiently early, so the universe is  
well inside the de Sitter phase. On the other hand, in mLQC-II and LQC, the BD vacuum can be imposed in the contracting phase as long as $t_0$ is sufficiently early, so
the universe becomes so large that the spacetime curvature is very small, and particle creation is negligible. With these in mind, the power spectra obtained in the three models, mLQC-I, mLQC-II and LQC, within the framework of the dressed metric approach were calculated and compared by imposing the initial conditions in the contracting phase. In particular, the spectra can be universally divided into three regimes, the infrared, intermediate and UV.
In the infrared and intermediate regimes, the relative differences in the amplitudes of the spectra can be as large as $100\%$ between mLQC-I and mLQC-II (the same is also true between mLQC-I  and LQC), but
in the UV regime such differences get dramatically reduced, which is no larger than $0.1\%$. Since the modes in the UV regime are the relevant ones to the current observations and also their corresponding power spectra are scale-invariant,  so these three models are all consistent with observations. 

In Sec. \ref{SecIV}, the cosmological perturbations of mLQC-I and mLQC-II are studied within the hybrid approach, and the subtleties of  the initial conditions are shown in Figs.  \ref{fig8}, \ref{fig9} and \ref{fig10},
 where Figs.  \ref{fig8} and \ref{fig9} are respectively for the quadratic and Starobinsky potentials in mLQC-I, while Fig.  \ref{fig10} is for the Starobinsky potential in mLQC-II. 
The case with the quadratic   potential  in mLQC-II is similar to that of mLQC-I, given by Fig.  \ref{fig8}.
 From these figures it is clear that  imposing the initial conditions now becomes a more delicate issue, and sensitively depends on the potential $V(\phi)$ of the inflaton field.  First, in the cases described by Figs. \ref{fig8} and \ref{fig9}, all the modes are oscillating during the time $ t_i^p < t < t_i$, so one might intend to impose   the BD vacuum at the bounce. However, for $t < t_i^p$ the quantity $\Omega_{\text{tot}}$ defined in Eq.(\ref{5.6}) experiences 
a period during which it is very large and negative. As a result, particle creation is  expected not to be negligible during this period. Then, imposing  {the BD vacuum} at the bounce will not account for these effects, and the resulting power spectra shall
be quite different from the case,  in which   in the deep contracting phase ($t \ll t_B$) the BD vacuum is imposed for mLQC-II and LQC, and  the de Sitter state for mLQC-I. On the other hand,  in the case described by Fig. \ref{fig10}, even if  the BD vacuum is chosen at the bounce,
it may not be quite different from the one imposed in the deep contracting phase, as now  in the whole contracting phase all the modes are oscillating, and 
particle creation is not expected  to be important up to the bounce.
To compare the results from the three different models, in this section the second-order adiabatic vacuum conditions are chosen in the contracting phase, which is expected not to be much different  from the de Sitter state   for mLQC-I and the BD vacuum for mLQC-II and LQC, as long as $t_0 \ll t_B$ in all the cases described by Figs. \ref{fig8} - \ref{fig10}.
The ambiguities of the choice of $\pi_a^{-2}$ and $\pi_a^{-1}$ also occur in this approach,  but as far as the power spectra are concerned, different choices lead to similar conclusions \cite{gqm2020}. 
So, in this section only the so-called prescription A is considered. Then, similar conclusions are obtained in this approach regarding the differences among the amplitudes of the power spectra in the three different models.
In particular, the relative differences  can be as large as $100\%$ between mLQC-I and mLQC-II/LQC, but in the UV regime such differences are reduced to about $2\%$. A remarkable feature between the two different approaches is also identified:  in the infrared and  oscillatory regimes, the power spectrum in mLQC-I is suppressed as compared with its counterpart in LQC  in the hybrid approach. On the other hand, in the dressed metric approach, the power spectrum in mLQC-I is largely amplified in the  infrared regime where its magnitude is as large as of the Planck scale \cite{IA19,lsw2020}. The main reason for such differences  is that
the effective mass in the hybrid approach is strictly positive near the bounce, while it is  strictly negative in the dressed metric approach for states that are initially dominated by the kinetic energy of the inflaton
\cite{IA19,lsw2020,gqm2020}. 

The review is concluded in Sec. \ref{SecV}, in which we summarize the main conclusions and point out some open questions for future studies.

\section{Effective Quantum Dynamics  in Modified LQCs}  
\label{SecII}
\renewcommand{\theequation}{2.\arabic{equation}}\setcounter{equation}{0}

To facilitate our following discussions,   let us first briefly review the standard process of quantization carried out in LQC, from which one can see clearly the similarities and differences among the three models, LQC, 
mLQC-I and mLQC-II.

\subsection{Quantum Dynamics of LQC}

The key idea of LQC is to use the  fundamental variables and quantization techniques of LQG to cosmological spacetimes, by taking  advantage of the simplifications that 
arise from the symmetries of these spacetimes. In the spatially-flat FLRW spacetime, 
\bqn
\lb{2.1}
ds^2&=&  -N^2(t)dt^2+ q_{ab}(t)dx^adx^b\nb\\
&\equiv&-N^2(t)dt^2+a^2(t)\delta_{ab} dx^adx^b ,  
\eqn
there exists only one degree of freedom, the scale factor $a(t)$, where $N(t)$ is the lapse function and can be freely chosen, given the freedom in reparametrizing $t$,
and $q_{ab}(t)$ denotes the 3-dimensional  (3D) spatial metric  of the hypersurface $t = $ Constant.  In this paper,  we shall use  the indices  $a, b, c, ...$ to denote spatial coordinates and $i, j, k, ...$ to denote the internal su(2) indices.  Repeated indices will represent  sum,   unless otherwise specified. 
 
In full GR, the gravitational phase space consists of the connection $A^i_a$ and  density weighted triad $E^a_i$. In the present case, 
the 3D spatial space $M$  has a ${\mathbb{R}}^3$ topology, from which we can introduce a fiducial cell $\mathcal{V}$ and restrict all integrations to this cell, in order to avoid some 
 artificial divergences and have a well-defined symplectic structure. Within this cell, we introduce a fiducial flat metric $\mathring{q}_{ab}$ via the relation $q_{ab}(t) = a^2(t) \; \mathring{q}_{ab}$, and then an associated constant orthogonal triad $\mathring{e}^a_i$ and 
 a cotriad $\mathring{\omega}^i_a$. Then, after symmetry reduction  $A^i_a$  and $E^a_i$ are given by, 
 \bqn
\lb{2.2}
A^i_a= c \; v_o^{-1/3} \; \mathring{\omega}^i_a,  \; E^a_i = |p| \; v_o^{-2/3} \sqrt{\mathring{q}} \; \mathring{e}^a_i,  
\eqn
where  $|p|= v_o^{2/3} a^2,\; \kappa = 8\pi G/c^4$, $v_o$ denotes the volume  of the fiducial cell measured by $\mathring{q}_{ab}$,  
$\mathring{q}$ is the determinant of $\mathring{q}_{ab}$, and $\gamma$  is the Barbero-Immirzi parameter whose value can be set to $\gamma \approx 0.2375$ using  black hole thermodynamics in LQG \cite{Mei04}. For classical solutions, symmetry reduced connection $c$ is related to time derivative of scale factor as $c = \gamma \dot a$, where an  over dot denotes a derivative with respect to $t$ for the choice $N = 1$. 

The physical triad and cotriad are given by $e_i^a = {\mbox{(sgn p)}} \; |p|^{-1/2} v_o^{1/3} \; \mathring{e}^a_i$ and $\omega^i_a = {\mbox{(sgn p)}}\; |p|^{1/2} v_o^{-1/3} \; \mathring{\omega}^i_a$, where ${\mbox{(sgn p)}}$ 
arises because in connection dynamics the phase space contains triads with both
orientations. In the following we choose this orientation to be positive and volume of the fiducial cell to be $v_o$ = 1. The variables $c$ and $p$ satisfy the communication relation,
 \bqn
\lb{2.3}
 \left\{c, p\right\} = \frac{\kappa\gamma}{3}.
\eqn
 {Then,} the gravitational part of the Hamiltonian   is a sum of the Euclidean  and Lorentzian   terms, 
\bq
\lb{2.4}
\mathcal{H}_{\mathrm{grav}} = {\cal H}_{\mathrm{grav}}^{(E)} - (1 + \gamma^2) {\cal H}_{\mathrm{grav}}^{(L)}, 
\eq
where, with the choice $N = 1$,   these two terms are given, respectively, by
\bqn
\lb{2.5a} 
\mathcal{H}_{\mathrm{grav}}^{(E)} &=& \frac{1}{2\kappa} \int \mathrm{d}^3 x \, \epsilon_{ijk} F^i_{ab} \frac{E^{aj} E^{bk}}{\sqrt{q}}, \\
\lb{2.5b}
\mathcal{H}_{\mathrm{grav}}^{(L)} &=&   \frac{1}{\kappa} \int \mathrm{d}^3 x \,   K^j_{[a} K^k_{b]}   \frac{E^{aj} E^{bk}}{\sqrt{q}}, 
\eqn
where $q= \mathrm{det}(q_{ab}) =    a^6 \; \mathring{q}$,  $F_{ab}^k$ is the field strength of  the connection $A^i_a$, and $K^i_a$  is the extrinsic curvature, given, respectively, by
\bqn
\lb{2.6}
F_{ab}^k &\equiv& 2\partial_{[a}A^k_{b]} + \epsilon_{ij}^{\;\;\;k}A^i_aA^j_b = c^2 \epsilon_{ij}^{\;\;\;k} \; \mathring{\omega}^i_a\;  \mathring{\omega}^j_b, \nb\\
K_a^i &\equiv& K_{ab} e^b_i = \frac{{e^b_i}}{2N} \left(\dot{q}_{ab} - 2D_{(a}N_{b)}\right) = \pm \dot{a} \; \mathring{\omega}^i_a. ~~~~
\eqn

 Upon quantization, ambiguities can arise  { due to different treatments of the Euclidean  and Lorentzian} terms in the Hamiltonian constraint.  
  In particular, LQC takes the advantage that in the spatially-flat FLRW universe the  Lorentzian part is proportional to the Euclidean part, 
  \bq
  \lb{2.7}
   {\cal H}_{\mathrm{grav}}^{(L)} = \gamma^{-2}{\cal H}_{\mathrm{grav}}^{(E)}, 
   \eq
 so that, when coupled to a massless scalar field, the classical Hamiltonian can be rewritten as \cite{abl,aps3} 
 \bqn
 \lb{2.8}
\mathcal{H}  &\equiv&   \mathcal{H}_{\mathrm{grav}} + \mathcal{H}_{M} \nb\\
&=&  - \frac{1}{2\kappa\gamma^2} \int \mathrm{d}^3 x \, \epsilon_{ijk} F^i_{ab} \frac{E^{aj} E^{bk}}{\sqrt{q}} + \mathcal{H}_{M}, 
\eqn 
where  $ \mathcal{H}_{M} = {p_{\phi}^2}/\left({2\sqrt{q}}\right)$, with $p_{\phi}$ being the momentum conjugate of $\phi$.

The  elementary operators in  the standard LQC are the triads\footnote{For a modification of LQC based on using gauge-covariant fluxes, see \cite{liegener-singh}.} $p$ and elements of the holonomies given by 
$\widehat{e^{i\bar{\mu} c/2}}$ of $c$, where $\bar{\mu} = \sqrt{\Delta  l_{pl}^2 /|p|}$ with $\Delta \equiv 4\sqrt{3}\pi \gamma$, and $\Delta  l_{pl}^2 $  being the minimum non-zero eigenvalue of the area operator, and the Planck length $l_{pl}$ is defined as
$l_{pl} \equiv \sqrt{\hbar G}$. However, it is found that, instead of using the eigenket $\left|p\right>$ of the area operator $p$ as the basis, it is more convenient to use the eigenket $\left|v\right>$ of the volume operator $\hat{v} \left(\equiv {\mbox{sgn}} (p)|p|^{3/2}\right)$,  
where
 \bq
 \lb{2.9}
 \hat{v} \left| v\right> = \left(\frac{8\pi \gamma}{6}\right)^{3/2}  \frac{|v|}{K} l_{pl}^3 \left| v\right>,\quad
 \widehat{e^{i\bar{\mu} c/2}} \left| v\right> = \left| v + 1\right>,  
\eq
with  $K \equiv 2\sqrt{2}\left(3\sqrt{3\sqrt{3}}\right)^{-1}$. Let $\Psi(v, \phi)$ denote the wavefunction  in the kinematical Hilbert space of the gravitational field coupled with the scalar field $\phi$, we have 
 \bqn
 \lb{2.10}
 \hat{\phi}\Psi(v, \phi) &=& \phi \Psi(v, \phi), \nb\\
  \hat{p}_{\phi} \Psi(v, \phi) &=& - i\hbar \frac{\partial}{\partial\phi} \Psi(v, \phi), \nb\\
  \widehat{|p|^{-3/2}}\Psi(v, \phi) &=& {\cal{B}}(v) \Psi(v, \phi),   
 \eqn
 where 
 \bqn
 \lb{2.11}
 {\cal{B}}(v) &\equiv& \left(\frac{6}{8\pi \gamma l_{pl}^2}\right)^{3/2}B(v),\nb\\
 B(v) &\equiv& \left(\frac{3}{2}\right)^{3}K |v|\left||v + 1|^{1/3} - |v-1|^{1/3}\right|^3. ~~~~
 \eqn
  Then,   the equation satisfied by selecting the physical states $\hat{\cal{H}}\Psi(v, \phi) = 0$ can be cast in the form,
  \bqn
 \lb{2.12}
\partial^2_{\phi}\Psi(v, \phi) &=& \frac{1}{B(v)} \Big[C^{+}(v) \Psi(v+4, \phi) - C^{o}(v) \Psi(v, \phi)\nb\\
&&  ~~~~~~~~~~ + C^{-}(v) \Psi(v-4, \phi)\Big],    
 \eqn
where 
\bqn
\lb{2.13}
C^{+}(v) &\equiv& \frac{3\pi K G}{8} |v+2|\left||v+1| - |v+3|\right|,\nb\\
C^{-}(v) &\equiv& C^{+}(v - 4), \;\;
C^{o}(v) \equiv C^{+}(v)+ C^{-}(v). ~~~~~~~
\eqn
This is the main result of LQC \cite{aps3}, which shows that: (i) It is {\em a  second order 
quantum difference equation with uniform discreteness in volume}, rather than a    { simple differential equation,  a direct consequence of the discrete nature of loop quantum geometry.
 (ii) It provides the evolution of the quantum cosmological wavefunction $\Psi(v, \phi)$, in which {\em the scalar field serves as a  clock}.  Thus, once an initial state  $\Psi(v, \phi_0)$ 
is given at the initial moment $\phi_0$, the study of the quantum dynamics of LQC can be carried out. It is found that, instead of a big bang  singularity,   a quantum bounce is generically produced, }
a result confirmed through extensive numerical simulations \cite{numlsu-1,numlsu-2,numlsu-3,numlsu-4,numlsu-5} and an exactly solvable model \cite{acs2010}. Using this model, one can compute the probability for the quantum bounce which turns out to be unity for an arbitrary superposition of wavefunctions \cite{craig-singh}.

For the states sharply peaked around a classical solution,  we can obtain ``effective" Friedmann and Raychaudhuri  {(FR)} equations, by using the geometric quantum mechanics 
in terms of the expectation values  of the operators $(\hat b, \hat v, \hat \phi, \widehat{p_{\phi}})$,  
 \bqn
 \lb{2.17a}
 \dot{b} &=& \left\{b, {{\cal{H}}}\right\}, \; \dot{v} = \left\{v, {{\cal{H}}}\right\}, \\
  \lb{2.17b}
  \dot{\phi} &=& \left\{\phi, {{\cal{H}}}\right\}, \;  \dot{p}_{\phi} = \left\{p_{\phi}, {{\cal{H}}}\right\},
  \eqn
which take the same forms as their classical ones, but all the quantities now  represent their expectation values, $A_I \equiv \left<\widehat{{A}_I}\right>$. 
Then, it was found that the effective Hamiltonian is given by \cite{aps3},
\bqn
\lb{2.17c}
{{\cal{H}}}_{\text{eff.}}   =  - \frac{3}{8\pi \gamma^2 \bar\mu^2 G}|p|^{1/2}\sin^2(\bar\mu c)  + \frac{1}{2}|p|^{3/2} p_{\phi}^2,
\eqn
which can also be expressed in terms of $v$ and $b$ via the relations $v=|p|^{3/2}$ and $b=c/\sqrt{|p|}$. Then, from  Eqs. (\ref{2.17a}) and (\ref{2.17b}) one can find that the ``effective" FR equations are given by,
 \bqn
 \lb{2.18a}
 H^2 &=&  \frac{8\pi G}{3}\rho\left(1 - \frac{\rho}{\rho_c}\right), \\
  \lb{2.18b}
 \dot{H} &=& -4\pi G \left(\rho + P\right) \left(1 - \frac{\rho}{\rho_c}\right),
 \eqn
where 
\bqn
\lb{2.19}
H &\equiv& \frac{\dot{v}}{3v} = \frac{\dot{a}}{a}, \quad  \rho_c \equiv \frac{3}{8\pi \lambda a^2\gamma^2 G}, \nb\\
\rho &\equiv& \frac{\mathcal{H}_M}{v}, \quad
P \equiv -\frac{\partial \mathcal{H}_M}{\partial v},
\eqn
and $v = v_o a^3$.    
 Since $H^2$ cannot be negative, from Eq.(\ref{2.18a}) we can see that we must have $\rho \le \rho_{c}$, and  at $\rho = \rho_{c}$ we have $H^2 = 0$,
that is, a quantum bounce  occurs at this moment. When  $\rho \ll \rho_{c}$,  the quantum gravity effects are negligible, whereby the classical relativistic limit is obtained.

For a scalar field $\phi$ with its potential $V(\phi)$, we have
\bq
\lb{2.19a}
{\mathcal{H}_M} \equiv  {\mathcal{H}_{\phi}}  =   v\left[\frac{p_{\phi}^2}{2v^2} +  V(\phi)\right].
\eq 
Then, from Eq.(\ref{2.17b}) we find that 
 \bqn
\lb{2.19ba}
\dot\phi &=& \frac{p_{\phi}}{v},\\
\lb{2.19bb}
\dot{p}_{\phi} &=& - v V_{,\phi}(\phi).  
\eqn

In the rest of  this review, we shall consider only the  states that are sharply peaked around a classical solution, so the above ``effective" descriptions are valid, and the 
questions raised recently  in \cite{KKL20} are avoided.

\subsection{Effective Dynamics of mLQC-I}  

As mentioned in the introduction,    an important open issue in LQC  is its connection with LQG   \cite{engle}. In particular,    in LQC the spacetime symmetry is first imposed (in the classical level), before
the quantization process is carried out. However, it is well-known that this is different from the general process of LQG \cite{LQG}, and as a result,  different Hamiltonian constraints could be resulted,
 hence resulting in  different Planck scale physics. 
Though the question of ambiguities in obtaining the Hamiltonian  in LQG is still open,   based on some rigorous proposals by Thiemann  \cite{thiemann},
various attempts have been carried out, in order  to obtain deeper insights into the question.

One of the first attempts to understand this issue was made in \cite{YDM09}, in which
 the Euclidean and Lorentzian terms given by Eqs.(\ref{2.5a}) and (\ref{2.5b}) are treated differently, by closely following the actual construction of LQG. To be more specific, in the full theory \cite{LQG}, the extrinsic curvature in the Lorentzian term (\ref{2.5b}) can be expressed in terms of the connection and the volume as 
 \bq
 K^i_a=\frac{1}{\kappa \gamma^3}\{A^i_a,\{\mathcal{H}_{\mathrm{grav}}^{(E)},V\}\},
 \eq
 which once substituted back into Eq. (\ref{2.5b}) lead to an expression of $\mathcal{H}_{\mathrm{grav}}^{(L)}$ different from that of $\mathcal{H}_{\mathrm{grav}}^{(E)}$ in the standard LQC (see \cite{YDM09} for more details).  Correspondingly, 
  one is able to obtain  the following ``effective" Hamiltonian   \cite{YDM09,lsw2018}, 
 \bqn
\lb{3.1}
\mathcal {H}^{\scriptscriptstyle{\mathrm{I}}}_{\text{eff.}}&=&\frac{3v}{8\pi G\lambda^2}\Big\{\sin^2(\lambda b)-\frac{(\gamma^2+1)\sin^2(2\lambda b)}{4\gamma^2}\Big\}\nb\\
&& ~~~~~~~~~~~~ +\mathcal{H}_M.
\eqn 
Hence,  the Hamilton's equations   take the  form,
\bqn
\lb{3.2a}
\dot v&=& 
\frac{3v\sin(2\lambda b)}{2\gamma \lambda}\Big\{(\gamma^2+1)\cos(2\lambda b)-\gamma^2\Big\}, ~~~\\
\lb{3.2b}
\dot b&=&  
\frac{3\sin^2(\lambda b)}{2\gamma \lambda^2}\Big\{\gamma^2\sin^2(\lambda b)-\cos^2(\lambda b)\Big\}\nb\\
&&  -4\pi G\gamma P,
\eqn
where $P$ represents the pressure defined in Eq.(\ref{2.19}).  Once the matter Hamiltonian $\mathcal{H}_M$ is specified, together with the
Hamiltonian constraint,
\bq
\lb{3.3}
\mathcal{H} \approx 0,
\eq
 Eqs.(\ref{3.2a}) and (\ref{3.2b}) uniquely determine the evolution of the universe. Using the non-graph changing
  regularization of the Hamiltonian \cite{thiemann},    the expectation values of the Hamiltonian operator yield 
the same ``effective" Hamiltonian of Eq.(\ref{3.1}) to the leading order \cite{DL17}.

It has been shown in detail that the big bang singularity is generically replaced by a quantum bounce when the energy reaches its maximum $ \rho_c^{\scriptscriptstyle{\mathrm{I}}}$ \cite{YDM09,DL17,lsw2018,lsw2018b,lsw2019}, 
where
\bqn
\lb{3.3a}
&&  \rho_c^{\scriptscriptstyle{\mathrm{I}}} \equiv \frac{\rho_c}{4(1+\gamma^2)},
 \eqn
 and the universe is asymmetric with respect to the bounce, in contrast to LQC.

To write Eqs.(\ref{3.2a})-(\ref{3.3}) in terms of $H,\; \rho$ and $P$, it was found that one must distinguish the pre- and post- bounce phases \cite{lsw2018}. In particular, before the bounce, 
the modified FR equations take the form  \cite{lsw2018},
\begin{widetext}
\bqn
\lb{3.4a}
H^2 &=&\frac{8\pi G\alpha  \rho_\Lambda}{3}\left(1-\frac{\rho}{\rho_c^{\scriptscriptstyle{\mathrm{I}}}}\right)\left[1+\left(\frac{1-2\gamma^2+\sqrt{1-\rho/\rho_c^{\scriptscriptstyle{\mathrm{I}}}}}{4\gamma^2\left(1+\sqrt{1-\rho/\rho_c^{\scriptscriptstyle{\mathrm{I}}}}\right)}\right)\frac{\rho}{\rho_c^{\scriptscriptstyle{\mathrm{I}}}}\right], \\
\lb{3.4b}
\frac{\ddot a}{a} &=&  - \frac{4\pi \alpha G}{3}\left(\rho + 3P - 2\rho_\Lambda \right) + 4\pi G\alpha P\left(\frac{2-3\gamma^2 +2\sqrt{1-\rho/\rho^{\scriptscriptstyle{\mathrm{I}}}_c}}{(1-5\gamma^2)\left(1+\sqrt{1-\rho/\rho^{\scriptscriptstyle{\mathrm{I}}}_c}\right)}\right)\frac{\rho}{\rho^{\scriptscriptstyle{\mathrm{I}}}_c} \nb\\
&& - \frac{4\pi \alpha G \rho}{3}\left[\frac{2\gamma^2+5\gamma^2\left(1+\sqrt{1-\rho/\rho^{\scriptscriptstyle{\mathrm{I}}}_c}\right)-4\left(1+\sqrt{1- \rho/\rho^{\scriptscriptstyle{\mathrm{I}}}_c}\right)^2}{(1-5\gamma^2)\left(1+\sqrt{1-\rho/\rho^{\scriptscriptstyle{\mathrm{I}}}_c}\right)^2}\right]\frac{\rho}{\rho^{\scriptscriptstyle{\mathrm{I}}}_c},
\eqn
\end{widetext}
where 
\bqn
\lb{3.5}
&& \alpha \equiv \frac{1-5\gamma^2}{\gamma^2+1}, \quad   \rho_\Lambda \equiv \frac{\gamma^2 \rho_c}{ (1+\gamma^2)(1-5\gamma^2)}.
\eqn

As $\rho \ll \rho_c^{\scriptscriptstyle{\mathrm{I}}}$,  Eqs.(\ref{3.4a}) and (\ref{3.4b}) reduce, respectively, to
\bqn
\lb{3.6a}
H^2&\approx&\frac{8\pi \alpha G}{3}\left(\rho+\rho_\Lambda\right), \\
\lb{3.6b}
\frac{\ddot a}{a}&\approx&-\frac{4\pi \alpha G}{3}\left(\rho+3P - 2\rho_\Lambda\right).
\eqn
 {These are exactly the FR  equations in GR for an ordinary matter field coupled with a  positive cosmological constant $\rho_\Lambda$, and a modified Newton's constant, $G_{\alpha} \equiv \alpha G$. 
For  $\gamma \approx 0.2375$, we have $\rho_\Lambda \approx 0.03  \rho_{pl}$, which is of the same order as  the one deduced conventionally in quantum field theory} 
 for the vacuum energy in our universe. In addition, we also have
 \bq
\lb{3.7}
\left|\frac{G_{{\alpha}}}{G} - 1\right|_{\gamma \approx 0.2375} \simeq 0.32 >  \frac{1}{8}.
 \eq
Finally, we want to emphasize that the minimal energy density of this branch, for which the Hubble rate vanishes, turns out to be negative which can be shown as $\rho_\mathrm{min}=-\frac{3}{8\pi G\lambda^2}\approx-0.023$. As a result, the necessary condition to generate a cyclic universe in mLQC-I is the violation of the weak energy condition which is in contrast to the cyclic universes in LQC where the energy density is always non-negative \cite{ls2021}.

In the post-bounce phase ($t > t_B$), from Eqs.(\ref{3.2a})-(\ref{3.3}) we find that \cite{lsw2018},
\begin{widetext}
\bqn
\lb{3.8a}
H^2 &=&\frac{8\pi G \rho}{3}\left(1-\frac{\rho}{\rho_c^{\scriptscriptstyle{\mathrm{I}}}}\right)\Bigg[1  +\frac{\gamma^2}{\gamma^2+1}\left(\frac{\sqrt{\rho/\rho_c^{\scriptscriptstyle{\mathrm{I}}}}}{1 +\sqrt{1-\rho/\rho_c^{\scriptscriptstyle{\mathrm{I}}}}}\right)^2\Bigg], \\
\lb{3.8b}
\frac{\ddot a}{a} &=&-\frac{4\pi G}{3}\left(\rho + 3P\right)
  + \frac{4\pi G \rho}{3}\left[\frac{\left(7\gamma^2+ 8\right) -4\rho/\rho_c^{\scriptscriptstyle{\mathrm{I}}} +\left(5\gamma^2 
14
 +8\right)\sqrt{1-\rho/\rho_c^{\scriptscriptstyle{\mathrm{I}}}}}{(\gamma^2 +1)\left(1+\sqrt{1-\rho/\rho_c^{\scriptscriptstyle{\mathrm{I}}}}\right)^2}\right]\frac{\rho}{\rho_c^{\scriptscriptstyle{\mathrm{I}}}}\nb\\
  &&  + 4\pi G P \left[\frac{3\gamma^2+2+2\sqrt{1-\rho/\rho_c^{\scriptscriptstyle{\mathrm{I}}}}}{(\gamma^2+1)\left(1+\sqrt{1-\rho/\rho_c^{\scriptscriptstyle{\mathrm{I}}}}\right)}\right]\frac{\rho}{\rho_c^{\scriptscriptstyle{\mathrm{I}}}},
\eqn
\end{widetext}
from which we obtain
\bqn
\lb{3.9}
\dot H&=&\frac{4 G \pi (P+\rho)}{(1+\gamma^2)}\left(2\gamma^2+2\frac{\rho}{\rho_c^{\scriptscriptstyle{\mathrm{I}}}}-3\gamma^2\sqrt{1-\frac{\rho}{\rho_c^{\scriptscriptstyle{\mathrm{I}}}}}-1\right).\nb\\
\eqn
Therefore, regardless of the matter content, the super-inflation  (starting at the bounce) will always end at $\rho = \rho_s$,   where 
\bq
\rho_s=\frac{\rho_c^{\scriptscriptstyle{\mathrm{I}}}}{8}\left(4-8\gamma^2-9\gamma^4+3\gamma^2\sqrt{8+16\gamma^2+9\gamma^4}\right),
\eq
for which we have $\dot H(\rho_s)=0$.

In the classical limit $\rho/\rho_c^{\scriptscriptstyle{\mathrm{I}}} \ll 1$, Eqs. (\ref{3.8a}) and (\ref{3.8b}) reduce, respectively, to
\bqn
\lb{3.9a}
H^2 &\approx& \frac{8\pi G}{3}\rho, \\
\lb{3.9b}
\frac{\ddot a}{a}&\approx&-\frac{4\pi G }{3}\left(\rho+3P\right),  
\eqn
whereby  the standard relativistic  cosmology is recovered. 

It is remarkable to note that in the pre-bounce phase the limit $\rho/\rho_c^{\scriptscriptstyle{\mathrm{I}}} \ll 1$ leads to Eqs.(\ref{3.6a}) and (\ref{3.6b}) with a modified Newtonian constant $G_{\alpha}$, 
while in the post-bounce the same limits leads to Eqs.(\ref{3.9a}) and (\ref{3.9b}) but now with the precise Newtonian constant $G$.

\subsection{Effective Dynamics of mLQC-II}
 
  In LQG,  the fundamental variables for the gravitational sector are the  su(2) Ashtekar-Barbero connection $A^i_a$ and  the conjugate triad $E^a_i$. When
  the Gauss and spatial diffeomorphism constraints are fixed, in the homogeneous  and isotropic universe the only relevant constraint is the Hamiltonian constraint, from which
  we obtain the FR equations,  as shown in the previous section. The Hamiltonian in mLQC-II arises from the substitution 
  \bq
  \lb{3.10}
  K^i_a= \frac{A^i_a}{\gamma},
  \eq
  in the Lorentzian term (\ref{2.5b}). 
 Then, the following effective Hamiltonian is resulted \cite{YDM09},
 \bqn
\lb{3.11}
\mathcal H^{\scriptscriptstyle{\mathrm{II}}}_\mathrm{eff.}&=&-\frac{3v}{2\pi G\lambda^2\gamma^2}\sin^2\left(\frac{\lambda b}{2}\right)\left\{1+\gamma^2\sin^2\left(\frac{\lambda b}{2}\right)\right\}\nb\\
&& +\mathcal{H}_M,
\eqn
from which we find that the corresponding  Hamilton's equations are given by, 
\bqn
\lb{3.12a}
\dot v&=&  \frac{3v\sin(\lambda b)}{\gamma \lambda}\Big\{1+\gamma^2-\gamma^2\cos\left(\lambda b\right)\Big\},\\
\lb{3.12b}
\dot b &=& -\frac{6\sin^2\left(\frac{\lambda b}{2}\right)}{\gamma \lambda^2}\Big\{1+\gamma^2\sin^2\left(\frac{\lambda b}{2}\right)\Big\}-4\pi G\gamma P\nb\\
&=& -4\pi G\gamma (\rho+P).
\eqn
 
 It can be shown that   the corresponding (modified) FR equations now read \cite{lsw2018b}, 
\bqn
\lb{3.13a}
H^2
&=&\frac{8\pi G \rho}{3}\left(1+\gamma^2 \frac{\rho}{\rho_c}\right) \left(1-\frac{(\gamma^2+1)\rho}{\Delta^2\rho_c}\right), ~~~\\
\lb{3.13b}
\frac{\ddot a}{a}
&=&-\frac{4\pi G}{3}\left(\rho+3P\right)-\frac{4\pi G P\rho}{\Delta \rho_c} \left[3\left(\gamma^2+1\right)-2\Delta\right]\nb\\
&&-\frac{4\pi G \rho^2}{3\Delta^2\rho_c}\left[7\gamma^2-1 +\left(5\gamma^2-3\right)\left(\Delta-1\right) -\frac{4\gamma^2\rho}{\rho_c}\right],\nb\\
\eqn
where $\Delta \equiv 1+\sqrt{1+\gamma^2\rho/\rho_c}$. 
  From these equations we can see  that  now the quantum bounce occurs when $\rho =  \rho_c^{\scriptscriptstyle{\mathrm{II}}}$,
 at which we have $H=0$ and $\ddot a >0$, where
\bq
\lb{3.14}
 \rho_c^{\scriptscriptstyle{\mathrm{II}}}=4(\gamma^2+1)\rho_c,
\eq
which is different  from the critical  density $\rho_c$ in LQC  as well as the  one $\rho_c^{{\scriptscriptstyle{\mathrm{I}}}}$  in mLQC-I.
Therefore,   the big bang singularity is also resolved in this model, and replaced by a  quantum bounce at $\rho = \rho_c^{\scriptscriptstyle{\mathrm{II}}}$,  similar to LQC and mLQC-I,
despite the fact that  the bounce in each of these models occurs at a different energy density. However, 
 in contrast to  mLQC-I, the evolution of the universe is symmetric with respect to the bounce, 
which is quite similar to the standard LQC model. 

 In addition,  similar to the other two cases, now the bounce is accompanied by a phase of super-inflation, i.e. $\dot H > 0$, which ends
at $\dot{H}(\rho_s) = 0$, but now $ \rho_s$ is given by,   
\bq
\lb{3.15}
\rho_s=\frac{\rho_c}{8 \gamma^2}\left(3(\gamma^2+1)\sqrt{1+2\gamma^2+9\gamma^4}+9\gamma^4+10\gamma^2-3\right).
\eq
For $\gamma=0.2375$, we find   $\rho_s=0.5132 \rho_c^{\scriptscriptstyle{\mathrm{II}}}$.  

When $\rho \ll \rho_c^{\scriptscriptstyle{\mathrm{II}}}$, the modified FR equations (\ref{3.13a})-(\ref{3.13b}) reduce  to,  
\bqn
\lb{3.16a}
H^2 &\approx& \frac{8\pi G}{3}\rho,\\
\lb{3.16b}
\frac{\ddot a}{a}&\approx&-\frac{4\pi G }{3}\left(\rho+3P\right),
\eqn
which are identical to those given in GR. Therefore, in this model, the classical limit is obtained in both pre- and post-bounce when  the energy density $\rho$ is much lower than  
the critical one $\rho_c^{\scriptscriptstyle{\mathrm{II}}}$.

\subsection{Universal Properties of mLQC-I/II Models}

 To study further  the evolution of the universe, it is necessary to specify the matter content $\mathcal H_M$. For a single scalar field with its  potential $V(\phi)$,
 the corresponding Hamiltonian takes  the  form (\ref{2.19a}).
As a result, the Hamilton's equations of the matter sector  are given by Eqs.(\ref{2.19ba}) and (\ref{2.19bb}).

The effective quantum dynamics of  LQC, mLQC-I, and mLQC-II were studied in detail recently in \cite{lsw2018b} for a single scalar field with various potentials, including 
the chaotic inflation, Starobinsky inflation, fractional monodromy inflation, non-minimal Higgs inflation, and inflation with an exponential potential, by using dynamical system analysis. 
 {It was found that, while several features of LQC were  shared by the mLQC-I and mLQC-II models, others belong to particular models.    In particular,
in the pre-bounce phase, the qualitative dynamics of LQC and mLQC-II  are quite similar, but  are strikingly different from that of mLQC-I. In all the three models, the non-perturbative quantum gravitational effects always 
result in a non-singular post-bounce phase, in which a short period of super-inflation always exists right after the bounce, and is succeeded by the conventional inflation.} The latter  is an  attractor in the phase space   for all the three models.

\begin{figure}[h!]  
{
\includegraphics[width=7cm]{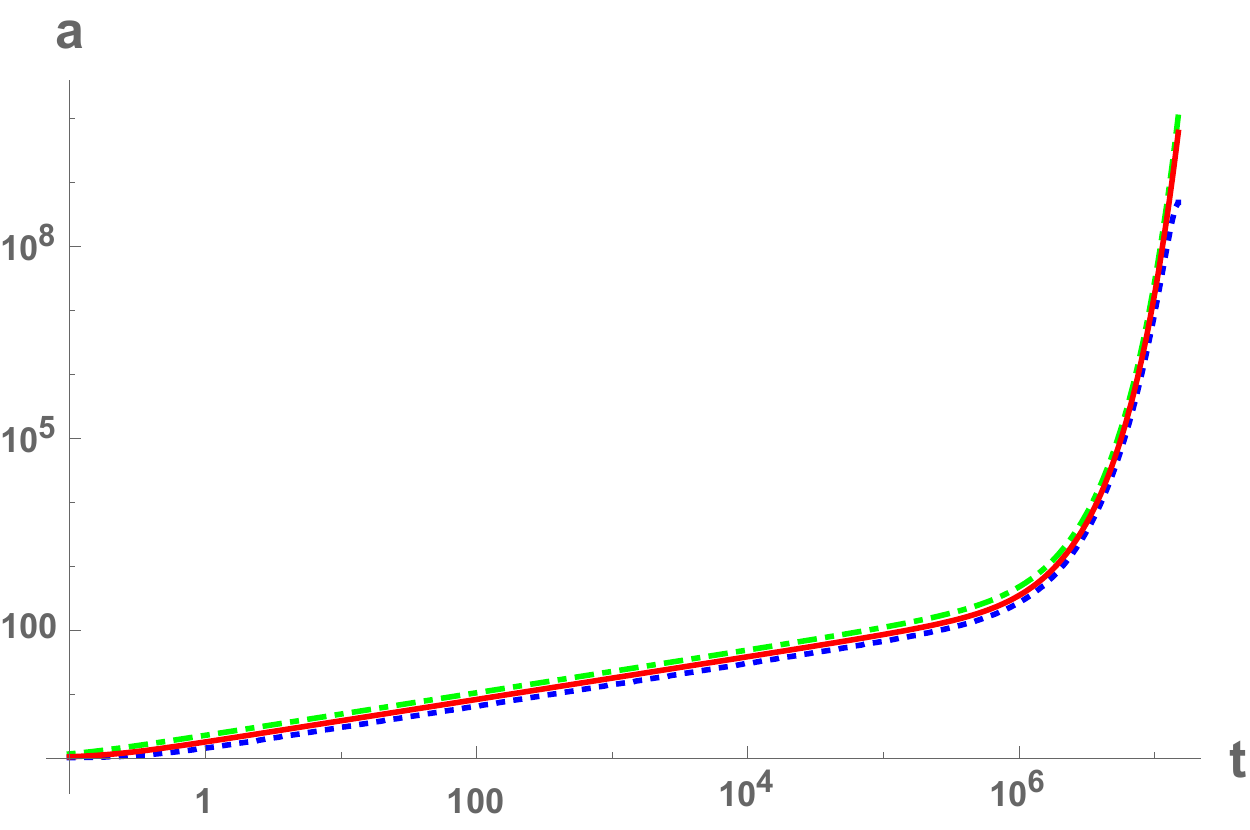}
\includegraphics[width=7cm]{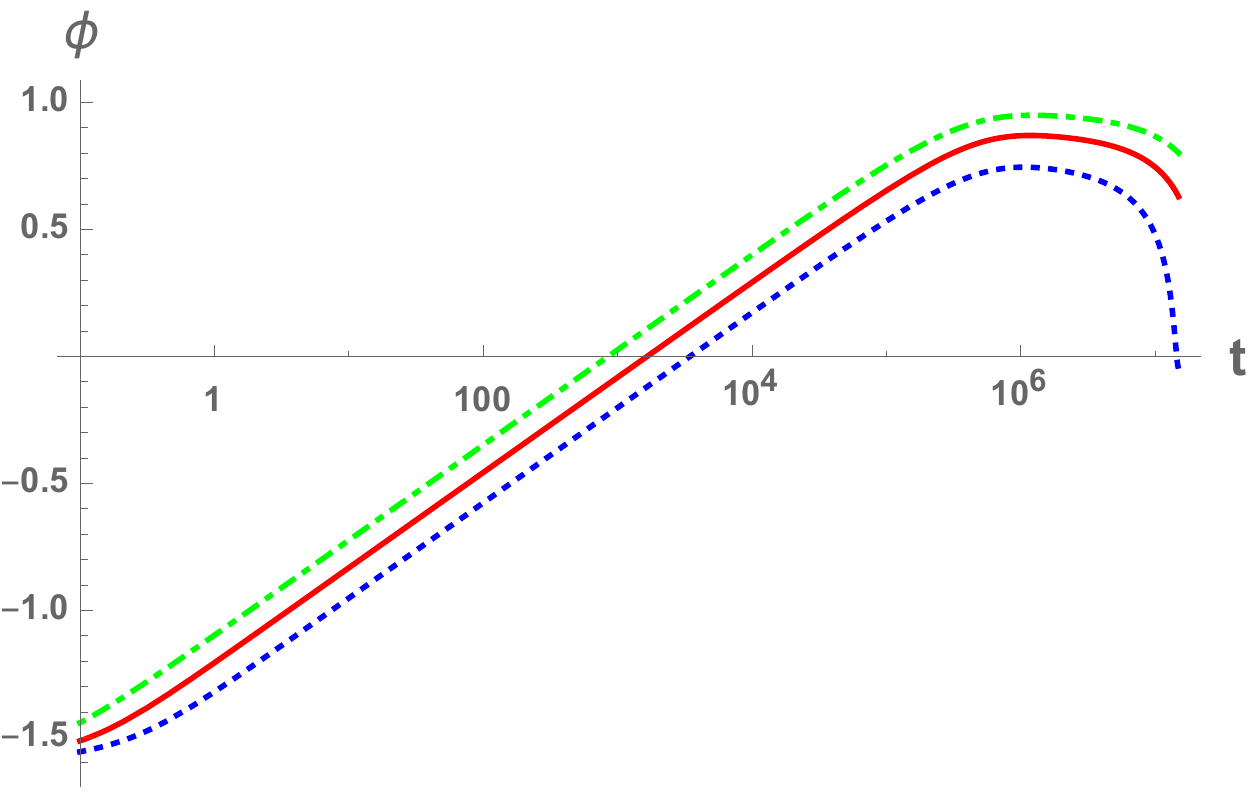}
\includegraphics[width=7cm]{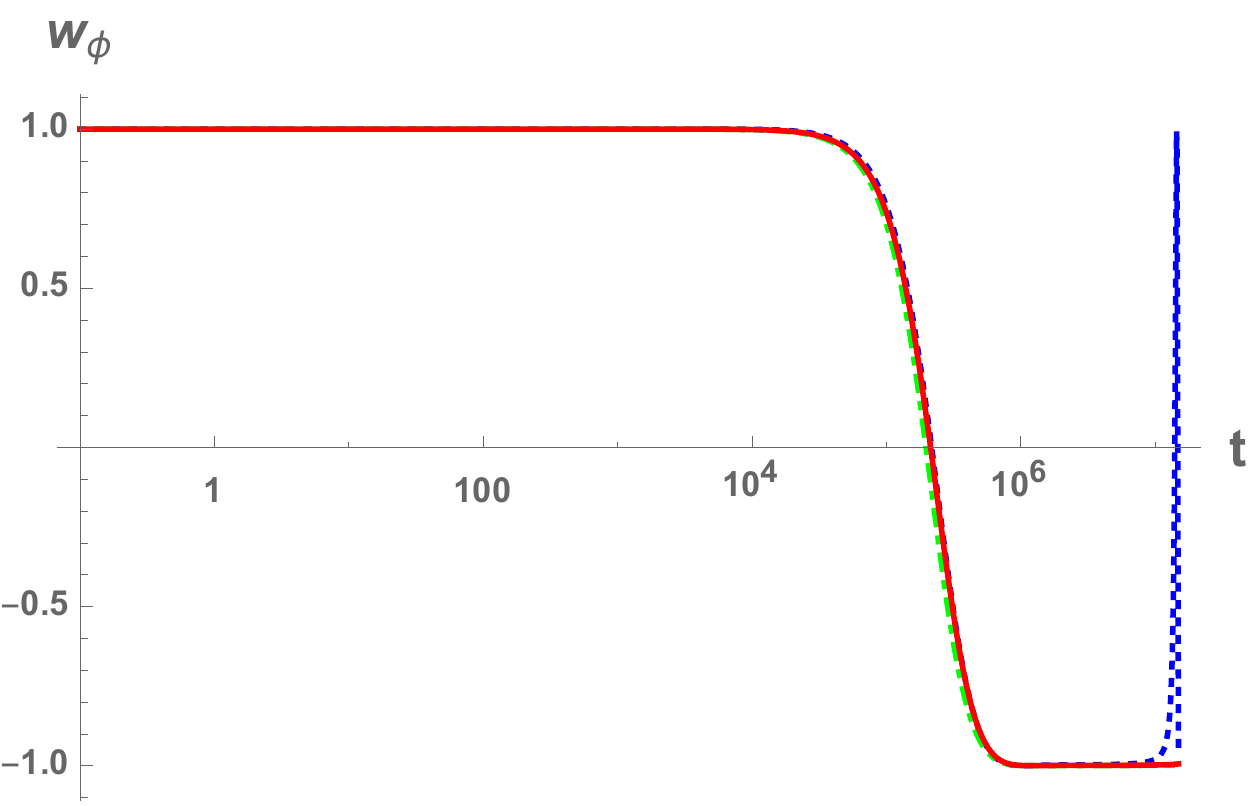}
}
\caption{The evolution of the scale factor $a(t)$, the scalar field $\phi(t)$, and the equation $w_{\phi}$ of state of the scalar field (from top to bottom) in the post-bounce phase are depicted and compared  among
the three modes,  LQC (red solid curves), mLQC-I (blue dotted curves) and mLQC-II (green dot-dashed curves), with the Starobinsky potential. In the last panel, $w_\phi$ is defined via ${w}_\phi\equiv P(\phi)/\rho(\phi)
= [\dot\phi^2 - 2V(\phi)]/[\dot\phi^2 + 2V(\phi)]$. The initial condition for the simulation is chosen at the bounce with  $\phi_B=-1.6~m_{pl} , \dot \phi_B>0$ \cite{lsw2019}.}
\label{fig1}
\end{figure}

 Moreover, similar to LQC  \cite{ZhuA,ZhuB} \footnote{In LQC, this universality was first found for the quadratic and Starobinsky potentials  \cite{ZhuA,ZhuB} (see also \cite{BCL19}), 
 and later was shown that  they are  also true for other potentials, including the  power-law potentials \cite{SSWW17,Shahalam18}, $\alpha$-attractor potentials  \cite{SSW18,SAMW20},  Monodromy potentials \cite{SSWW18}, warm inflation \cite{XW20}, Tachyonic inflation \cite{KX20a} and even in Brans-Dicke LQC \cite{SZW19,JMZ19}.},  for the initially kinetic energy dominated  conditions,
 \bq
 \lb{3.19}
 \frac{1}{2} \dot{\phi}_B^2 \gg V(\phi_B),
 \eq
 it was found that the evolution of the universe before the reheating is universal. In particular, in the post-bounce phase (between the quantum bounce and the reheating), the evolution can be uniquely divided into three phases: {\em bouncing, transition and slow-roll inflation}, as shown in Fig. \ref{fig1} for the Starobinsky potential,
 \bq
 \lb{3.20}
V(\phi) =\frac{3m^2}{32\pi G}\left(1-e^{-\sqrt{16\pi G/3} \phi}\right)^2.
\eq
For other potentials, similar results can be obtained, as long as at the bounce  the evolution of the universe is dominated by  the kinetic energy of the inflaton $w(\phi_B) \simeq 1$ \cite{lsw2019,KX20b}. 
 
In each of these three phases, the evolutions of $a(t)$ and $\phi(t)$ can be well approximated by  analytical solutions.  In particular, during the bouncing phase, they are given by 
\bqn
\lb{3.21}
a(t)&=&\left[1+ 24 \pi G  \rho_c^{\scriptscriptstyle{\mathrm{I}}} \left(1+\frac{A \gamma^2}{1+B t}\right)t^2\right]^{1/6},\nb\\
\phi(t)&=& \phi_{\text{B}} \pm \frac{m_{pl} \; \text{arcsinh}{\left(\sqrt{24\pi G   \rho_c^{\scriptscriptstyle{\mathrm{I}}} \left(1+\frac{C \gamma^2}{1+Dt}\right) }t\right)}}{\sqrt{12\pi G \left(1+\frac{C \gamma^2}{1+D t}\right)}},\nb\\
\eqn
for mLQC-I model, where the parameters $A$, $B$, $C$ and $D$ are fixed through numerical simulations. It was found that 
the best fitting   is provided by \cite{lsw2019},
\bqn
\lb{3.22}
A=C=1.2, \quad B=6, \quad D=2. 
\eqn

For the mLQC-II model, during the bouncing phase $a(t)$ and $\phi(t)$ are given by 
\bqn
\lb{3.23}
a(t)&=&\left[1+ 24 \pi G \rho_c^{\scriptscriptstyle{\mathrm{II}}} \left(1+\frac{A \gamma^2}{1+B t}\right)t^2\right]^{1/6},\nb\\
\phi(t)&=& \phi_{\text{B}} \pm \frac{m_{pl} \; \text{arcsinh}{\left(\sqrt{24\pi G  \rho_c^{\scriptscriptstyle{\mathrm{II}}} \left(1+\frac{C \gamma^2}{1+Dt}\right) }t\right)}}{\sqrt{12\pi G \left(1+\frac{C \gamma^2}{1+D t}\right)}},\nb\\
\eqn
 but now with,  
\bq
\lb{3.24}
A=2.5, \quad B=7, \quad C=D=2.
\eq 

In the transition and slow-roll inflationary phases, the functions  $a(t)$ and $\phi(t)$ were  given explicitly in \cite{lsw2019}. 

For the initially potential energy dominated  cases,
 \bq
 \lb{3.25}
 \frac{1}{2} \dot{\phi}_B^2 \ll V(\phi_B),
 \eq
 it was found that such universalities are lost. In particular, for the Starobinsky potential, the potential energy dominated bounce cannot  give rise to  any period of inflation for both
 mLQC-I and mLQC-II models, quite similar to what happens in LQC \cite{bonga_inflation_2016,bonga_phenomenological_2016}.

\subsection{Probabilities of the Slow-roll Inflation in mLQC-I/II Models}

 To consider the probability of the slow-roll inflation in the modified LQC models,   {let us start with  the phase space $\mathbb{S}$ of  the modified Friedmann and  Klein-Gordon equations,
 which now is four-dimensional (4D), and consists of the four variables,  $(v, \; b)$ and $(\phi, \; p_\phi)$,  from  the gravitational and matter sectors, respectively. 
 Using the canonical commutation relations, the symplectic form on the 4D phase space is given by \cite{SVV06,ZL07,as2011,CK11,corichi-sloan,bedic,LB13,CZ15},
\bq
\lb{4.1}
\Omega=d p_\phi \wedge d\phi+ \frac{d v \wedge d b}{4 \pi G \gamma}.
\eq 
However, after taking   the effective Hamiltonian constraint into account, 
\bq
\lb{4.2}
\mathcal{C}= 16\pi G \left\{{\cal{H}}_{grav}(v, b)  + \frac{p_{\phi}^2}{2v} + v V(\phi) \right\} \simeq 0,
\eq
where $``\simeq"$ means that the equality  holds only on $\bar\Gamma$, we can see that  the 4D phase space $\mathbb{S}$ reduces  to a three-dimensional (3D) hypersurface $\bar\Gamma$.}

 On the other hand,  the phase space $\mathbb{S}$ is isomorphic to a 2-dimensional (2D) gauge-fixed surface $\hat\Gamma$ of $\bar\Gamma$, which is intersected by each dynamical trajectory once and only once \cite{as2011}. From the FR equations, it can be shown that for both mLQC-I and mLQC-II the variable $b$ satisfies the equation \cite{lsw2018,lsw2018b,lsw2019},
\bqn
\lb{4.3}
\dot{b} = - 4\pi G \gamma(\rho + P).
\eqn
For any given matter field that satisfies the weak energy condition \cite{HE73}, we have $\rho + P > 0$, so the function $b$ is monotonically decreasing. Then, a natural parameterization of this 2D surface is $b = $ constant, say, $b_0$. Hence, using the   Hamiltonian constraint (\ref{4.2}) we find
\bq
\lb{4.3a}
p^{\mathrm{A}}_{\phi} = v \left\{-2\left[\hat{\cal{H}}_{grav}^{\mathrm{A}}   +  V(\phi)\right]\right\}^{1/2},
\eq
where $\mathrm{A} = \mathrm{I}, \mathrm{II}$, and 
\bq
\lb{4.4}
\hat{\cal{H}}_{grav}^{\mathrm{A}} \equiv v^{-1} {\cal{H}}_{grav}^{\mathrm{A}}(v, b_0).
\eq  
On the other hand, from Eqs.(\ref{3.1}) and (\ref{3.11}) we find that
$\hat{\cal{H}}_{grav}^{\mathrm{A}} = \hat{\cal{H}}_{grav}^{\mathrm{A}}(b_0) = $ constant on $\hat\Gamma$. Thus, we find 
\bqn
\lb{4.5}
\left. dp^{\mathrm{A}}_{\phi}\right|_{\hat\Gamma} =  \frac{p^{\mathrm{A}}_{\phi}}{v} dv - \frac{v^2 V_{,\phi}}{p_{\phi}} d\phi.  
\eqn
 Inserting this  expression into Eq.(\ref{4.1}), we find that the pulled-back symplectic structure $\hat\Omega$ reads
 \bqn
\lb{4.6}
\left. \hat\Omega^{\mathrm{A}}\right|_{\hat\Gamma} =   \left\{-2\left[\hat{\cal{H}}_{grav}^{\mathrm{A}}(b_0)   +  V(\phi)\right]\right\}^{1/2} d\phi \wedge dv,
\eqn
from which we find that  the Liouville measure $d\hat\mu_L$ on $\hat\Gamma$ is given by
\bqn
\lb{4.7}
d\hat\mu^{\mathrm{A}}_L =   \left\{-2\left[\hat{\cal{H}}_{grav}^{\mathrm{A}}(b_0)   +  V(\phi)\right]\right\}^{1/2} d\phi dv.  
\eqn

 Note that   $d\hat\mu^A_L$   does not depend on $v$, so that the integral with respect to $dv$ is infinite. However, this divergency shall be cancelled when calculating  the probability,
 as it will appear   in both denominator and numerator.   Therefore, the measure for the space of physically distinct solutions can be finally taken as
\bq
\lb{4.8}
d\omega^{\scriptscriptstyle{\mathrm{A}}}=\left\{-2\left[\hat{\cal{H}}_{grav}^{\mathrm{A}}(b_0)   +  V(\phi)\right]\right\}^{1/2} d\phi,
\eq
so that the 2D phase space $\hat\Gamma$ is further reduced to an interval  $\phi\in \left(\phi_{\mathrm{min}}, \phi_{\mathrm{max}}\right)$.  
It should be noted that   such a defined measure depends explicitly on $b_0$, and its choice in principle is arbitrary. However, in loop cosmology  there exists a preferred choice, which
is   its value   at the quantum bounce  $b_0 = b\left(t_B\right)$ \cite{as2011}.
With such a choice, the probability of the occurrence of an event $E$ becomes
\bq
\lb{4.9}
P(E) = \frac{1}{{\cal{D}}} \int_{\mathcal{I}(E)}{\left\{-2\left[\hat{\cal{H}}_{grav}^{\mathrm{A}}(b_B)   +  V(\phi)\right]\right\}^{1/2} d\phi},
\eq
where $\mathcal{I}(E)$ is the interval on the $\phi_B$-axis, which corresponds to the physically distinct initial conditions in which the event $E$ happens, and ${\cal{D}}$ is the total measure 
\bq
\lb{4.10}
 {\cal{D}} \equiv  \int_{\phi_{\mathrm{min}}}^{\phi_{\mathrm{max}}} {\left\{-2\left[\hat{\cal{H}}_{grav}^{\mathrm{A}}(b_B)   +  V(\phi)\right]\right\}^{1/2} d\phi}.
\eq

 Once the  probability is properly defined,  we can calculate it in different models of the modified LQCs.  In LQC \cite{as2011}, the calculations were carried out for the quadratic potential.
In order to compare the results obtained in different models, let us consider the same potential.  Then, for the mLQC-I model it was found that \cite{lsw2019}
\bqn
\lb{4.11}
&& \sin\left(\lambda b_B^{\scriptscriptstyle{\mathrm{I}}}\right)=\sqrt{\frac{1}{2\gamma^2+2}}, \quad \sin\left(2\lambda b_B^{\scriptscriptstyle{\mathrm{I}}}\right)=\frac{\sqrt{2\gamma^2+1}}{\gamma^2+1},\nb\\
&& p^{\scriptscriptstyle{\mathrm{I}}}_\phi=v\left(2\rho^{\scriptscriptstyle{\mathrm{I}}}_c-2 V\right)^{\frac{1}{2}}, \quad
d\omega^{\scriptscriptstyle{\mathrm{I}}}=\left(2\rho^{\scriptscriptstyle{\mathrm{I}}}_c-2 V\right)^{\frac{1}{2}} d\phi, ~~~~~~~~
\eqn
so that the probability for the desired slow-roll not to happen is,
\bq
\lb{4.13}
P^{\scriptscriptstyle{\mathrm{I}}}(\text{not realized})\lesssim\frac{\int^{0.917}_{-5.158} d \omega^{\scriptscriptstyle{\mathrm{I}}} }
{\int^{\phi^{\scriptscriptstyle{\mathrm{I}}}_{\text{max}}}_{-\phi^{\scriptscriptstyle{\mathrm{I}}}_{\text{max}}}{ d \omega^{\scriptscriptstyle{\mathrm{I}}}}} \simeq 1.12\times 10^{-5},
\eq
where $\phi^{\scriptscriptstyle{\mathrm{I}}}_{\text{max}} = 3.49\times 10^5\; m_{pl}$. 

 In mLQC-II, following a similar analysis, it can be shown that  the probability for the desired slow-roll not to  happen  is \cite{lsw2019}, 
\bq
\lb{4.14}
P^{\scriptscriptstyle{\mathrm{II}}}(\text{not realized})\lesssim 2.62\times 10^{-6}.
\eq
Note that in LQC the probability for the desired slow roll inflation not to occur is \cite{as2011},
\bq
\lb{4.15}
P^{\scriptscriptstyle{\mathrm{LQC}}}(\text{not realized})\lesssim2.74\times 10^{-6},
\eq
which is smaller  than that  for mLQC-I and slightly larger than the one for mLQC-II. However, it is clear that  the desired slow-roll inflation is very  likely to occur in all the models,
including the two  modified LQC ones.


\section{Primordial power spectra  of Modified LQCs in dressed metric approach}  
\label{SecIII}
\renewcommand{\theequation}{3.\arabic{equation}}\setcounter{equation}{0}

As mentioned above, in the literature there exist several approaches to investigate the inhomogeneities of the universe.
Such approaches can be generalized to the modified LQC models, including mLQC-I and mLQC-II. In this section we shall focus ourselves 
on cosmological  perturbations in the framework of mLQCs following the dressed metric approach \cite{aan2012,aan2013,aan2013b}, while in the next section we will be following the  hybrid approach
 \cite{mm2012, mm2013, gmmo2014,gbm2015,mo2016}. We shall  restrict ourselves to the effective dynamics, as we did with the homogeneous 
background in the last section. Such investigations in general include two parts: (a) the initial conditions; and (b) the dynamical evolutions of perturbations. In the framework of 
effective dynamics, the latter is a second-order ordinary differential equation in the momentum space, so in principle once the initial conditions are given, it uniquely determines
the cosmological (scalar and tensor) perturbations. 

However,  the   initial conditions   are a subtle issue,  which is true not only  in LQC but also in  mLQCs.
This is mainly because that in general there does not exist  a preferred initial time and state for a quantum field in an arbitrarily curved space-time \cite{BD82}.
If the universe is sufficiently   {smooth} and its evolution is  sufficiently slow,  so that the characteristic scale  of perturbations is much larger than the wavelengths
of all the relevant modes,  a well justified  initial
state can be defined: {\it the BD vacuum}. This is precisely the initial state commonly adopted in GR at the beginning of the slow-roll inflation,
in which  all the relevant perturbation modes are well inside {\it the
comoving Hubble radius} \cite{DB09} [cf. Fig.\ref{fig2}].

  However, in LQC/mLQCs,  especially near the bounce, the evolution of the background is far from ``slow", and its geometry is also far from the de Sitter. 
 In particular, for the perturbations during the bouncing phase,
 the wavelengths could be larger, equal, or smaller than the corresponding characteristic scale, as it can be seen, for example,  from Fig. \ref{fig5}. Thus, it is in general impossible to assume that the universe is in the
 BD vacuum     at the bounce \cite{aan2013b,d1,d1b,ZhuA,ZhuB}.   Therefore, in the following let us first elaborate in more details about  the subtle issues regarding  the   initial conditions.

\subsection{Initial Conditions for Cosmological Perturbations}

\begin{figure}[h!]  
{
\includegraphics[width=7cm]{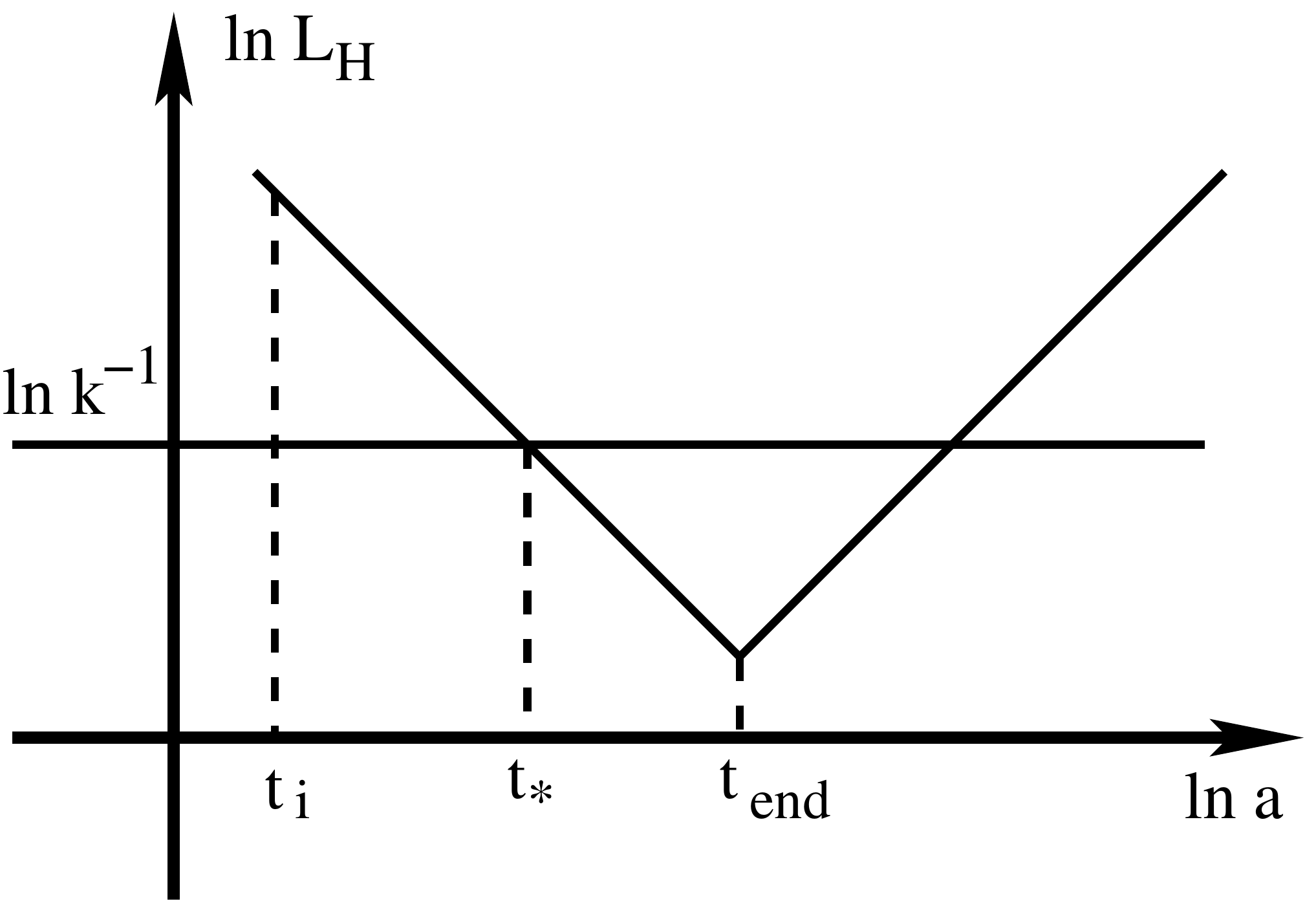}
}
\caption{The evolution of the comoving Hubble radius $\ln(L_H)$ vs $\ln(a)$ in GR, where $t_i$ denotes the moment of the onset of the slow-roll inflation, $t_*$ the horizon crossing time of a mode with the
 wavenumber $k$, and $t_{\text{end}}$ the moment that the slow-roll inflation ends.}
\label{fig2}
\end{figure}

 The initial conditions for cosmological perturbations in fact consists of two parts: when and which?   However, both questions are related to each other. In LQC literature, for cosmological perturbations,  two different moments have been chosen so far in the dressed and hybrid approaches:
 (i) the remote past in the contracting phase \cite{lsw2020}  and  (ii)  the bounce \cite{aan2013b,d1,d1b}. To see which conditions we need to impose at a given moment, let us first recall how to impose the initial conditions in GR, in which the
  scalar  perturbations are governed by the equation,
  \bqn
  \lb{5.1}
  v_k'' + \left(k^2 - \frac{z''}{z}\right)v_k = 0,
  \eqn
 where $k$ denotes the comoving wave number, and $z \equiv a\dot{\delta\phi}/H$, with $\delta\phi$ being the scalar field perturbations, $\phi = \bar\phi(t) + \delta\phi(t, x)$. A prime denotes a derivative with respect to the conformal time $\eta$, while an over dot denotes a derivative with respect to the cosmic time $t$, where $d\eta = dt/a(t)$.  The standard choice of the initial sate is the Minkowski vacuum of an  incoming observer in the far past, $k \gg aH$ [cf. Fig. \ref{fig2}]. In this limit, Eq.(\ref{5.1}) becomes $v_k'' + k^2 v_k = 0$, which has the solution,
  \bq
  \lb{5.3a}
  v_k  \simeq \frac{\alpha_k}{\sqrt{2k}}e^{-ik\eta} +  \frac{\beta_k}{\sqrt{2k}}e^{ik\eta},
  \eq
 where $\alpha_k$ and $\beta_k$ are two integration constants, and must satisfy the normalized condition, 
  \bq
  \lb{5.4}
  v_k^* v_k' - {v_k^*}'v_k = - i.
  \eq
  If we further require  {\it the vacuum to be the minimum energy state}, a unique solution exists,  which is given by $\alpha_k = 1, \; \beta_k = 0$, 
  that is,
   \bq
  \lb{5.3}
 \lim_{k \gg aH} v_k  \simeq \frac{1}{\sqrt{2k}}e^{-ik\eta},
  \eq
  which is often referred to as {\it the BD vacuum} \cite{DB09}.

 Consider the de Sitter space as the background, we have $a(\eta) = 1/(-\eta H)$, and
 $z''/z = a''/a = 2/L_H^2$, where $L_H \equiv 1/(aH) = -\eta$ is the corresponding  comoving Hubble radius. Then, Eq.(\ref{5.1}) reads, 
   \bqn
  \lb{5.2}
  v_k'' + \left(\frac{1}{\lambda^2} - \frac{2}{L_H^2}\right)v_k = 0,
  \eqn
  where $\lambda \; (\equiv 1/k)$ denotes the comoving wavelength.  The above equation has the following exact solutions,
    \bq
  \lb{5.2a}
  v_k =  \frac{\alpha_k}{\sqrt{2k}}e^{-ik\eta}\left(1 - \frac{i}{k\eta}\right)  +  \frac{\beta_k}{\sqrt{2k}}e^{ik\eta} \left(1 + \frac{i}{k\eta}\right). 
  \eq
   It is clear that on scales much smaller than the comoving Hubble radius  ($ \lambda \ll L_H$), $v_k$ is oscillating with frequency $k$ and constant amplitude, given by Eq.(\ref{5.3a}).  
   Then,  setting $(\alpha_k, \; \beta_k) = (1, 0)$ we find that Eq. (\ref{5.2a}) reduces to 
    \bq
  \lb{5.2b}
  v_k =  \frac{1}{\sqrt{2k}}e^{-ik\eta}\left(1 - \frac{i}{k\eta}\right). 
  \eq
 Note that if the initial time $t_i$ is chosen sufficiently small, i.e., $t_i \ll t_{\text{end}}$ or $|k\eta| \gg 1$, all the modes are inside the comoving Hubble radius $L_H$
 [cf. Fig. \ref{fig2}], 
 and the BD vacuum (\ref{5.3}) becomes a natural choice.
  
  However, on the scales  much larger than the comoving Hubble radius  ($ \lambda \gg L_H$), the $k^2$ term is negligible compared to the squeezing term, $z''/z$,
  and as a result, the fluctuations will stop oscillating and the amplitude of $v_k$ starts to increase,  yielding
   \bq
  \lb{5.5}
  v_k  \simeq z(\eta).
  \eq
 
  As shown in Fig. \ref{fig2}, if the initial time $t_i$ is chosen to be sufficiently early, all the currently observed modes $k_{\text{ph}} \in (0.1, 1000)\times k_0^*$ will be well inside the  {comoving}  Hubble radius
  at $t = t_i$, so the mode function $v_k$ can be well approximately given by Eq.(\ref{5.3}), which is the well-known zeroth order adiabatic state, where $k_0^* = 0.002$ Mpc$^{-1}$ and $k_{\text{ph}}(t) \equiv k/a(t)$ \cite{cosmo}.

In modified LQC models, the mode function $v_k$   satisfies the following modified equation,
 \bqn
  \lb{5.6}
  v_k'' + \Omega_{\text{tot}}^2(\eta, k) v_k = 0,
  \eqn
 where  $\Omega_{\text{tot}}^2(\eta, k)$ depends on  the homogeneous background  {and the inflation potential $V(\phi)$,
 so it is  model-dependent. Therefore,} the choice of the initial conditions will depend on not 
 only the modified LQC models to be considered but also the moment at which the  initial conditions  are imposed. 
 
 One of the main reasons to choose the remote past in the contracting phase as the initial time for
perturbations is that at such time either the background is well described by the de Sitter space
(mLQC-I) or the expansion factor $a(t)$ becomes so large that the curvature of the background is
negligible (mLQC-II and LQC),   so imposing the BD vacuum for mLQC-II and LQC and  the de Sitter state given by
Eq.(\ref{5.2b}) for mLQC-I at this moment  is well justified. It should be noted that the reason to refer  {to} the state described by 
Eq.(\ref{5.2b}) as  {\it the de Sitter state} is the following:  In  the slow-roll inflation, the homogeneous and isotropic universe is almost
de Sitter, as the Hubble parameter $H \equiv \dot{a}/a$ is almost a constant, so we have $a(\eta) \simeq 1/(-H\eta)$. For $t_i \ll t_{\text{end}}$ we have $a(\eta) \ll 1$, and
  $|\eta k| \simeq |H\eta| \gg 1$,
 so the choice $(\alpha_k \; \beta_k) = (1, 0)$ will lead Eq.(\ref{5.2b}) directly to Eq.(\ref{5.3})  at the onset of the slow-roll inflation [cf. Fig. \ref{fig2}]. However, 
 in the deep contracting phase of the same de Sitter space, now the universe is very large, that is, $a(\eta) \gg 1$, so we must have $|H\eta| \ll 1$
 and   $|\eta k| \simeq |k/H_{\Lambda}|a^{-1}(\eta)$, which can be very small in sufficiently  early times of the contracting phase, so the terms $i/(k\eta)$ 
 in Eq.(\ref{5.2a}) now cannot be neglected. To distinguish this case from the one described by Eq.(\ref{5.3}), 
  In this review we  refer the state described by Eq.(\ref{5.2b})  with the term $i/(k\eta)$ not being negligible as the  de Sitter state, while  {the state described by Eq.(\ref{5.3})  is still called}  the  BD vacuum state, or simply the BD vacuum.

   On the other hand, if the initial time is 
 chosen to be  at the bounce, cautions must be taken on what initial conditions can be imposed {\it consistently}. In particular, if  at this moment some modes are inside the comoving Hubble 
 radius and others are not,  it is clear that in this case  imposing  the BD vacuum at the bounce will lead to inconsistencies.
 In addition, there also exist the cases in which
  particle   creation  in the contracting phase is not negligible, then it is unclear how a BD vacuum  can  be imposed at the bounce, after the universe is contracting for such a long time before the bounce. 
  Thus, in these cases  other initial conditions need to be  considered, such as the fourth-order adiabatic  vacuum  \cite{aan2013b,d1,d1b,ZhuA,ZhuB}.

   With the above in mind, in the following we turn to consider power spectra of the cosmological perturbations.

 \subsection{Power Spectra of Cosmological Perturbations}
 
 Since the evolutions of  the effective dynamics of the homogeneous backgrounds  for mLQC-I and mLQC-II are different, in this subsection let us first consider the case of mLQC-I and then  mLQC-II.
  To compare the results with those obtained in LQC, at the end of this subsection, we also  {discuss} the LQC case.
 
 \subsubsection{mLQC-I}

 For mLQC-I, the power spectrum of the cosmological scalar perturbations was first studied in \cite{IA19}  for the quadratic $\phi^2$ potential,  and re-examined   in \cite{lsw2020}. 
 In the terminology used in \cite{IA19}, it was found that  the corresponding mode function $v_k (\equiv q_k/a)$ satisfies Eq.(\ref{5.6}) with
  \bqn
  \lb{5.7}
  \Omega_{\text{tot}}^2 &=& k^2 - \frac{a''}{a} + \mathfrak{A}_{-}, \quad r \equiv \frac{24\pi G \dot\phi^2}{\rho}, \nb\\
    \mathfrak{A}_{-} &\equiv& a^2\left[V(\phi) r - 2V_{,\phi}(\phi)\sqrt{r} + V_{,\phi\phi}(\phi)\right], 
  \eqn
  where $r=24\pi G\dot \phi^2/\rho$ and  $V(\phi)$ denotes the scalar field potential.

It should be noted that, when generalizing the classical expression of the function $\mathfrak{A}_{-}$ defined in Eq.(\ref{5.7}) to its corresponding quantum mechanics operator,  
there exists ambiguities. In fact, classically $\mathfrak{A}_{-}$ only coincides with  $\Omega^2_{Q}$ in the expanding phase.   The latter is a function of the phase space variables which is explicitly given by \cite{aan2013b,abs2018,IA19,lsw2020}, 
\bq
\lb{5.9}
 \Omega^2_{Q}=3\kappa  \frac{p^2_\phi}{a^4}-18\frac{p^4_\phi}{a^6\pi^2_a}-12a\frac{p_\phi}{\pi_a}V_{,\phi}+a^2V_{,\phi\phi},
\eq
where $\pi_a$ is the moment conjugate to $a$, and given by one of Hamilton's dynamical equations,
\bq
\lb{5.10}
\pi_a = - \frac{3a^2}{4\pi G} H,
\eq
with the choice of the lapse function $N = 1$. Therefore, $\pi_a < 0 \; (\pi_a > 0)$ corresponds to  $H > 0 \; (H < 0)$. At the quantum bounce we have $H(t_B) = 0$, 
so that $\pi_a (t_B) = 0$. Then, $ \Omega^2_Q$ defined by Eq.(\ref{5.9}) diverges at the bounce. Hence, from the Friedmann equation, $H^2 = (8\pi G/3) \rho$,  we find that 
\bq
\lb{5.11}
\frac{1}{\pi_a^2} = \frac{2\pi G}{3 a^4 \rho}, \quad \frac{1}{\pi_a} = \pm \sqrt{\frac{2\pi G}{3 a^4 \rho}}, 
\eq
where ``-" corresponds to $H > 0$, and ``+"   to $H < 0$. Then, a direct  generalization leads to \cite{lsw2020},   
\bq
\lb{5.11aa}
\Omega^2_{Q} = \begin{cases}
\mathfrak{A}_{-}, & H \ge 0,\cr
\mathfrak{A}_{+}, & H \le 0,\cr 
\end{cases}
\eq
where 
\bq
\lb{5.11a}
\mathfrak{A}_{\pm}  \equiv a^2\left[V(\phi) r \pm 2V_{,\phi}(\phi)\sqrt{r} + V_{,\phi\phi}(\phi)\right].
\eq
It should be noted  that in \cite{IA19} only the function   $\mathfrak{A}_{-}$ was chosen over the whole process of the evolution of the universe.   The same choice  was
also adopted in \cite{abs2018,AKS20}.

In addition,  $\mathfrak{A}$ defined by Eq.(\ref{5.11}) is not continuous across the bounce, as the coefficient $2V_{,\phi}(\phi)\sqrt{r}$ in general does not vanish at the bounce.
In \cite{ZhuA,ZhuB,nbm2018} 
$\mathfrak{A}_{-}$ appearing in Eq.(\ref{5.7}) was replaced by $U(\phi) (\equiv \Omega^2_{+})$ over the whole process of the evolution of the homogeneous universe, 
where
\bq
\lb{5.11c}
\Omega^2_{\pm} \equiv a^2\left[\mathfrak{f}^2 V(\phi)   \pm 2 \mathfrak{f} V_{,\phi}(\phi) + V_{,\phi\phi}(\phi)\right],
\eq
 and $\mathfrak{f} \equiv (24\pi G/\rho)^{1/2}\; \dot{\phi}$. 
 
\begin{figure}[h!] 
{
\includegraphics[width=8cm]{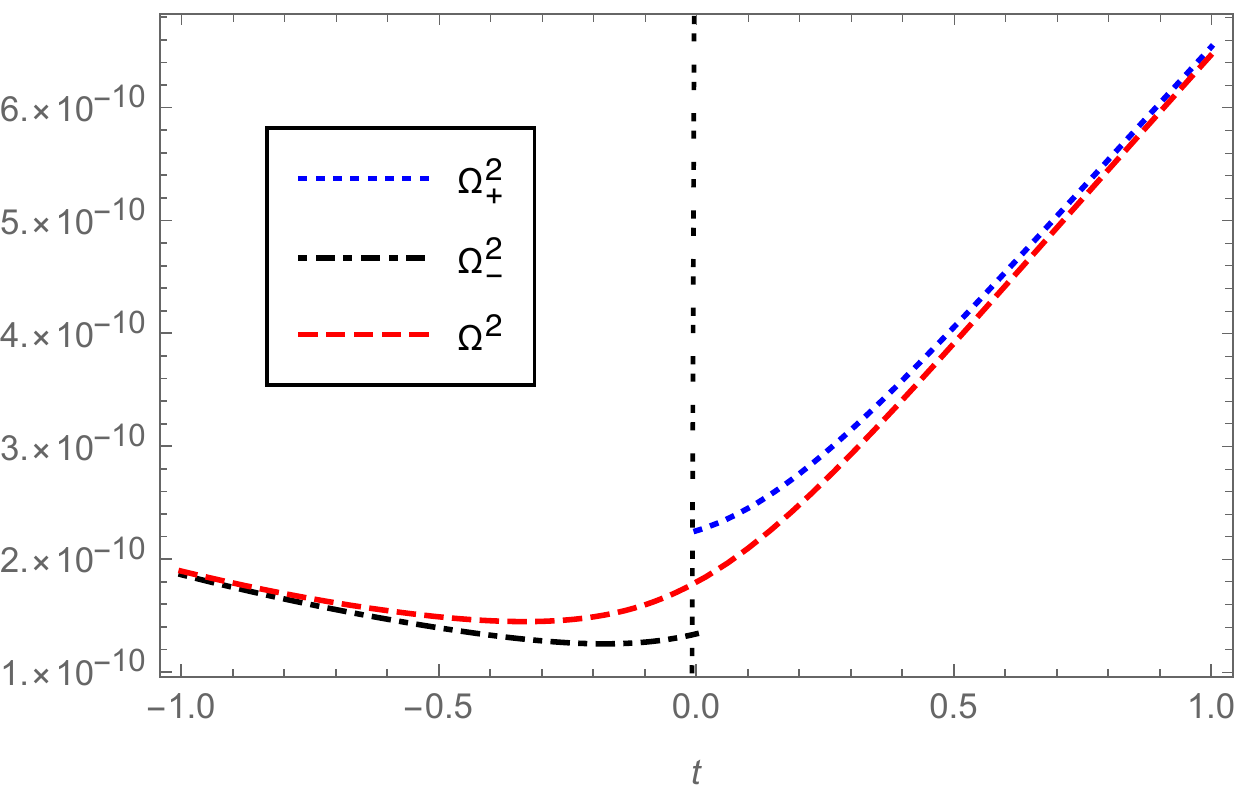}
}
\caption{The potential terms   $\Omega^2_+$ and $\Omega^2_-$ are compared with their smooth extension $\Omega^2$ across the bounce in mLQC-I  {for the quadratic potential $V(\phi) = 
\frac{1}{2} m^2\phi^2$} \cite{lsw2020}.
}
\label{fig3}
\end{figure}

Another choice was introduced  in \cite{lsw2020}, which was motivated from the following considerations. The functions $\Omega^2_{\pm}$ defined above
 are not continuous across the bounce, quite similar to $\mathfrak{A}_{\pm}$.  However, if we introduce the quantity 
  $\Omega^2$ as, 
\bq
\lb{5.11d}
\Omega^2=a^2 \left[\mathfrak{f}^2 V(\phi)   + 2  \Theta(b) \mathfrak{f} V_{,\phi}(\phi) + V_{,\phi\phi}(\phi)\right],  
\eq
to replace  $\mathfrak{A}_{-}$ in Eq.(\ref{5.7}), it could be continuous across the bounce by properly choosing
$\Theta(b)$. In particular, the variable $b(t)$ satisfies  {Eq.(\ref{4.3})} \cite{lsw2018,lsw2018b,lsw2019} \footnote{It is interesting to note that Eq.(\ref{4.3}) holds not only in 
mLQC-I, but also in LQC and mLQC-II.}, 
from which we can see that $b(t)$ is always a monotonically decreasing function  for any matter that satisfies the weak energy condition \cite{HE73}. 
Moreover, one can construct a step-like function of $b$ with the bounce as its symmetry axis \cite{lsw2018,lsw2018b,lsw2019}. Therefore,
if we define $ \Theta(b)$ as 
\bqn
\lb{5.12}
 \Theta(b) =  1-2\left(1+ \gamma^2\right)\sin^2\left(\lambda b\right),
\eqn
it behaves precisely as a step function, so that $\Omega^2$ smoothly connects $\Omega_{\pm}^2$ across the bounce, as shown in Fig. \ref{fig3}.

\begin{figure}[h!] 
{
\includegraphics[width=8cm]{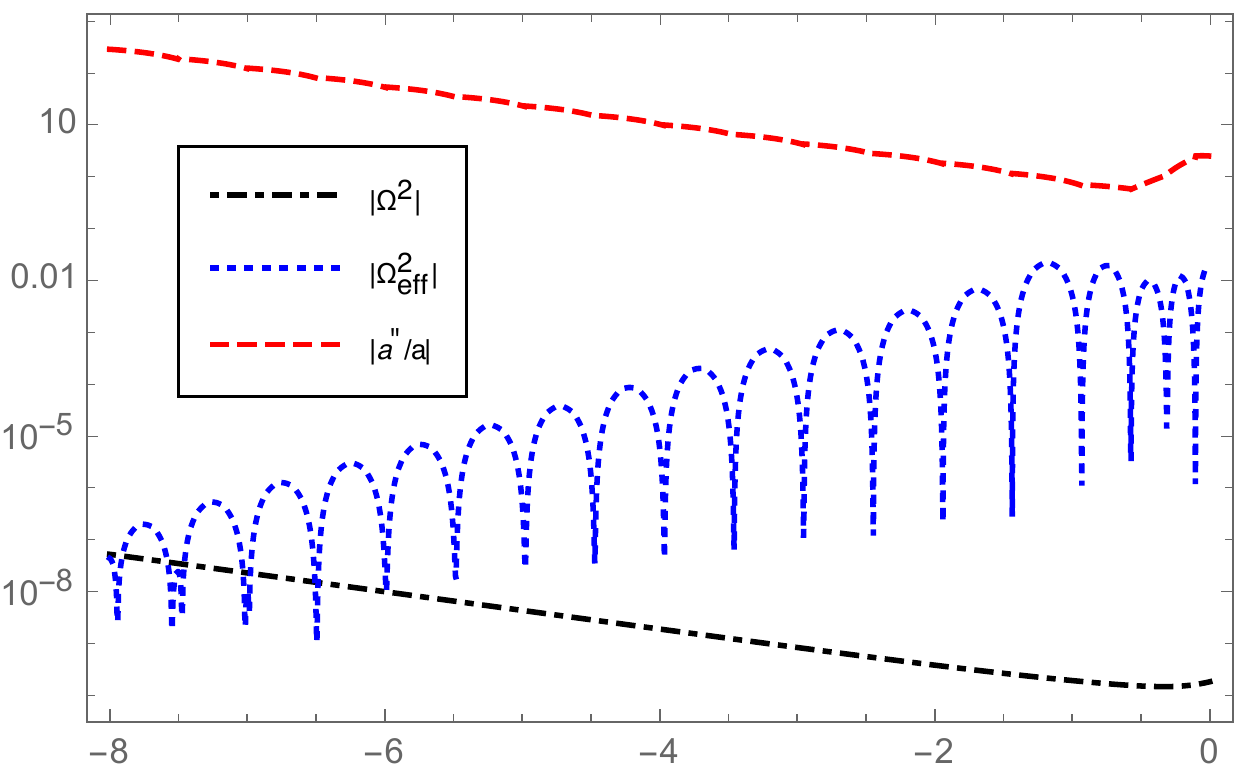}
\includegraphics[width=8cm]{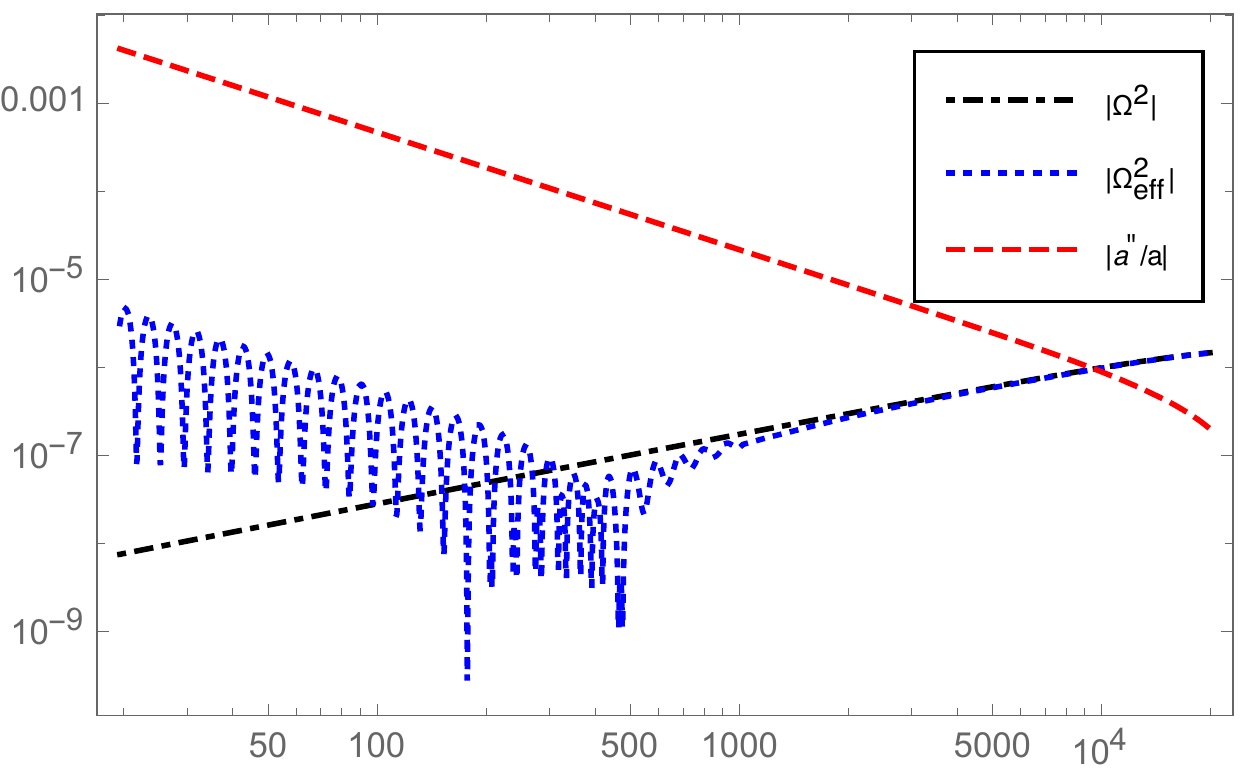}
}
\caption{The potential terms $\Omega^2$ and  $\Omega^2_\mathrm{eff}$ are compared with  the curvature term $a^{\prime \prime}/a$  in mLQC-I near the bounce and the preinflationary regime
  {for the quadratic potential $V(\phi) = \frac{1}{2} m^2\phi^2$}  \cite{lsw2020}.
}
\label{fig4}
\end{figure}

In addition to the above choices, motivated by the  hybrid approach  \cite{mm2012,mm2013,gmmo2014,gbm2015,mo2016}, 
 the following replacements for $\pi_a^{-2}$ and $\pi_a^{-1}$ in Eq.(\ref{5.9}) were also introduced in  \cite{lsw2020}, 
\bqn
\lb{5.13a}
\frac{1}{\pi^2_a}&\rightarrow& \frac{64 \pi^2 G^2\lambda^2 \gamma^2}{9a^4{\cal{D}}}, \\
\lb{5.13b}
\frac{1}{\pi_a}&\rightarrow& -\frac{8 \pi G \lambda  \gamma \Theta(b)}{3a^2{\cal{D}}^{1/2}},
\eqn
where
\bqn
\lb{5.14}
{\cal{D}} \equiv  \left(1+\gamma^2\right)\sin^2\left(2\lambda b\right)-4\gamma^2\sin^2\left(\lambda b\right). 
\eqn
Such obtained $\Omega^2_{Q}$ was referred to as $\Omega_{\text{eff}}^2$ in \cite{lsw2020}, and in Fig. \ref{fig4}, we show the differences among $\Omega^2$, $\Omega_{\text{eff}}^2$
and the quantity $a''/a$, from which one can see that the term $a''/a$ dominates  the other two terms over the whole range $t/ t_{pl} \in \left(-8, 10^4\right)$.

 To  study the effects of the  curvature term,  let us first introduce the quantity,  
\bq
\lb{5.15}
k_B^{\scriptscriptstyle{\mathrm{I}}} =\left. \left(\frac{a''}{a}\right)^{1/2}\right|_{t=t_B}\approx1.60,
\eq
which is much larger than other two terms $\Omega^2$ and $\Omega^2_\mathrm{eff}$,  where
$\Omega^2(t_B)=1.75\times 10^{-10}$ and  $\Omega^2_\mathrm{eff}(t_B) =0.006$.  Therefore, the differences between $\Omega^2$ and $ \Omega^2_\mathrm{eff}$ near the bounce are highly
diluted by the background.  On the other hand, in the post-bounce phase, $\Omega^2$ and $\Omega^2_\mathrm{eff}$ coincide after $t/t_{pl} \simeq10^4$, while near the onset of the inflation
 their amplitudes first become almost 
equal to that of  the curvature term, and then quickly exceeds it  during the slow-roll inflation, as we can see from Fig. \ref{fig4}.

 From Fig. \ref{fig4} we can also see that  the difference between $\Omega^2$ and $ \Omega^2_\mathrm{eff}$ lies mainly in the region near the bounce. However, as the curvature term $a''/a$ overwhelmingly  dominates in this region, it is  usually expected that the impact of the different choices of $\Omega^2$ on the power spectrum might not be very large \cite{aan2013b,abs2018,IA19}. However, 
in \cite{lsw2020}, it was found that the  relative difference in the magnitude of the power spectrum in the IR and oscillating regimes could be as large as  $10\%$, where the  relative difference is defined
as,
\bq
\lb{5.15a}
{\cal{E}} \equiv 2\left|\frac{{\cal{P}}_1 - {\cal{P}}_2}{{\cal{P}}_1 + {\cal{P}}_2}\right|. 
\eq

However,  the power spectra obtained from $\Omega^2$ and $\Omega^2_{\pm}$  are substantially different even in the UV regime due to the (tiny) difference between $\Omega^2_{\pm}$ at the bounce, see Fig. 14 given in  \cite{lsw2020}. In fact, the difference is so large that   the power spectrum calculated from $\Omega^2_{\pm}$ is essentially already ruled out by current observations.

   With the clarification of the ambiguities caused by the quantum mechanical generalization of the  function $\mathfrak{A}_{-}$ defined in Eq.(\ref{5.7}), now let us turn to the issue of the initial conditions,
  for which we consider only two representative potentials, the quadratic and Starobinsky,  given explicitly by
   \bq
 \lb{5.29a}
  V =\begin{cases}
  \frac{1}{2}  m^2\phi^2, & \text{quadratic},\cr
  \frac{3m^2}{32\pi G}\left(1 - e^{-\sqrt{16\pi G/3}\; \phi}\right)^2, & \text{Starobinsky}.\cr
  \end{cases}
  \eq

 In mLQC-I, the evolution of the effective (quantum) homogeneous universe is asymmetric with respect to the bounce \cite{DL17,lsw2018,lsw2018b,lsw2019}. In particular, before the bounce ($t < t_B$), the universe is asymptotically de Sitter,
 and only very near the bounce (about several Planck  {seconds}), the Hubble parameter $H$  {which is negative in the pre-bounce regime} suddenly increases to zero at the bounce. 
 Then, the universe enters a very short super-acceleration phase $\dot{H} > 0$ (super-inflation) right after the bounce, which lasts  until $\rho \simeq \rho_c^{\scriptscriptstyle{\mathrm{I}}}/2$, where $\dot{H}(t)_{\rho \simeq \frac{1}{2}\rho_c^{\scriptscriptstyle{\mathrm{I}}}} = 0$.
  Afterwards, for a kinetic energy dominated bounce $\dot\phi^2_B \gg V(\phi_B)$, it takes about  $10^{4} \sim 10^{6}$ Planck seconds before entering the slow-roll inflation  \cite{lsw2018,lsw2018b,lsw2019}. 
 Introducing the quantity,
  \bqn
  \lb{5.8}
  \lambda_H^2 \equiv \frac{a}{{a''} - a\mathfrak{A}_{-}} = - \frac{1}{m_{\text{eff}}},
  \eqn
where  
$m_{\text{eff}}$ is the effective mass of the modes, from Eq.(\ref{5.7}) we find that 
\bq
\lb{5.8aa}
\Omega_{\text{tot}}^2 = \frac{1}{\lambda^2} -  \frac{1}{\lambda_H^2} =
\begin{cases}
> 0, & \lambda_H^2 > \lambda^2,\cr
< 0, & 0 < \lambda_H^2 < \lambda^2,\cr
> 0, &  \lambda_H^2 < 0,\cr
\end{cases}
\eq
where $\lambda\; (\equiv k^{-1})$ denotes the comoving wavelength of the mode $k$,   as mentioned above. Note that such a defined quantity $\lambda_H^2$ becomes negative when  the effective mass
is positive.  
In Fig. \ref{fig5},  we plot $\lambda_H^2$  schematically for the quadratic and the Starobinsky potentials with the mass of the inflaton set to $1.21\times 10^{-6}\;m_\mathrm{pl}$ and $2.44\times 10^{-6}\;m_\mathrm{pl}$ respectively.  The initial conditions for the background evolution   are set as follows: for the quadratic potential, the inflaton starts with a positive velocity on the right wing of the potential and for the Starobinsky potential the inflaton is released from the left wing of the potential with a positive velocity. For both potentials, the initial conditions are set at the bounce which is dominated by the kinetic energy of the inflaton field. The same mass parameters and similar initial conditions are also used in the following figures where the comoving Hubble radius is plotted schematically. In Fig. \ref{fig5},  
the moments $t_H$ and $t_i$ are defined, respectively, 
by $a''(t_{H}) = a''(t_i)=0$, so
$t_i$ represents the beginning  of the inflationary phase,
and during the slow-roll inflation (Region III), we have $\lambda_H^2\approx L_\text{H}^2/2 \simeq 1/(2a^2H^2)$, which is exponentially decreasing, and all the  modes observed today were inside the comoving Hubble radius at $t=t_i$. Between the times $t_H$ and $t_i$, $\lambda_H^2$ is negative, and $\Omega^2_{\text{tot}}$   is strictly positive. Therefore,   during this period the mode functions are oscillating, while
during the epoch  between $t_B$ and $t_H$,  some modes ($k^{-2} > k_B^{-2}$) are inside the comoving Hubble radius,  and others ($k^{-2} < k_B^{-2}$) are outside it right after the bounce, where
$k_B\equiv \lambda_B^{-1}(t_B)$. 
 In the contracting phase,  {when $t \ll  t_B$, the universe is quasi-de Sitter  and  $\lambda_H^2 \simeq 1/(2a^2H^2)$ increases exponentially  toward  the bounce $t \rightarrow t_B$, as now $a(t)$ is decreasing exponentially. However, several Planck seconds before the bounce, the universe enters a non-de Sitter state, during which $\lambda_H^2$ starts to decrease until the bounce, at which  a characteristic Planck
 scale $k_B\; (\equiv 1/\lambda_H)$ can be well defined.
 Therefore,  for $t \ll t_B$, all the modes are outside the comoving Hubble radius. Then, following our previous arguments, if the initial moment is chosen at $t_0 \ll t_B$,   the de Sitter state    seems not  to be viable. However, when $t_0 \ll t_B$ we have $a(\eta) \simeq 1/(-\eta|H_{\Lambda}|)$, where $H_{\Lambda} = - [\lambda (1+\gamma^2)]^{-1}$ and 
 \bq
 \lb{5.8a} 
 \Omega_{\text{tot}}^2  \simeq k^2 -  \frac{2}{\eta^2}, 
 \eq
 for which Eq.(\ref{5.6}) has the exact solutions given by Eq.(\ref{5.2a}). 
 Therefore, at sufficient early times,  choosing $\alpha_k = 1, \; \beta_k = 0$ leads us to  the de Sitter state (\ref{5.2b}).  From the above analysis it is clear that this is possible 
 precisely because of the isometry of the de Sitter space, which  is sufficient to single out a preferred state,  the de Sitter state \cite{IA19}.   
 
 With the exact solution (\ref{5.2b}) as the initial conditions imposed at the moment $t_0 \; (\ll t_B)$ in the contracting phase, it was found that the power spectrum of the cosmological scalar perturbations can be divided into three different regimes: (i) the ultraviolet (UV)  ($k > k_{\text{mLQC-I}}$); (ii)  intermediate ($k_i < k < k_{\text{mLQC-I}}$); and (iii)  infrared  ($k < k_{i}$), where $k_{\text{mLQC-I}} \equiv a_B\sqrt{R_B/6}$ and
 $k_i = a_i \sqrt{R_i/6}$, and $R_B$ and $R_i$ are the curvatures given at the bounce and the beginning of the slow-roll inflation, respectively [cf. Fig. \ref{fig5}]. During the infrared regime, the power spectrum increases as $k$ increases, 
 while in the  intermediate regime it is oscillating very fast and the averaged amplitude of the power spectrum is decreasing as $k$ increases. In the UV regime,  the spectrum is almost scale-invariant, which is consistent with the current observations. There exists a narrow band, $0.1 \times k_0^* < k < k_{\text{mLQC-I}}$, in which the quantum gravitational effects could be detectable by current or forthcoming cosmological observations \cite{IA19}. Within the dressed metric approach, one of the most distinctive features of the power spectrum in mLQC-I is that its magnitude in the IR regime  is of the Planck scale \cite{IA19,lsw2020}. This is because those infrared modes are originally outside the Hubble horizon in the contracting phase and thus their magnitudes are frozen as they propagate across the bounce and then into the inflationary phase. Considering that the contracting phase is a quasi de Sitter phase with a Plank-scale Hubble rate, the magnitude of the IR modes is thus also Planckian \cite{lsw2020}. 
  
\begin{figure}[h!]  
{
\includegraphics[width=7cm]{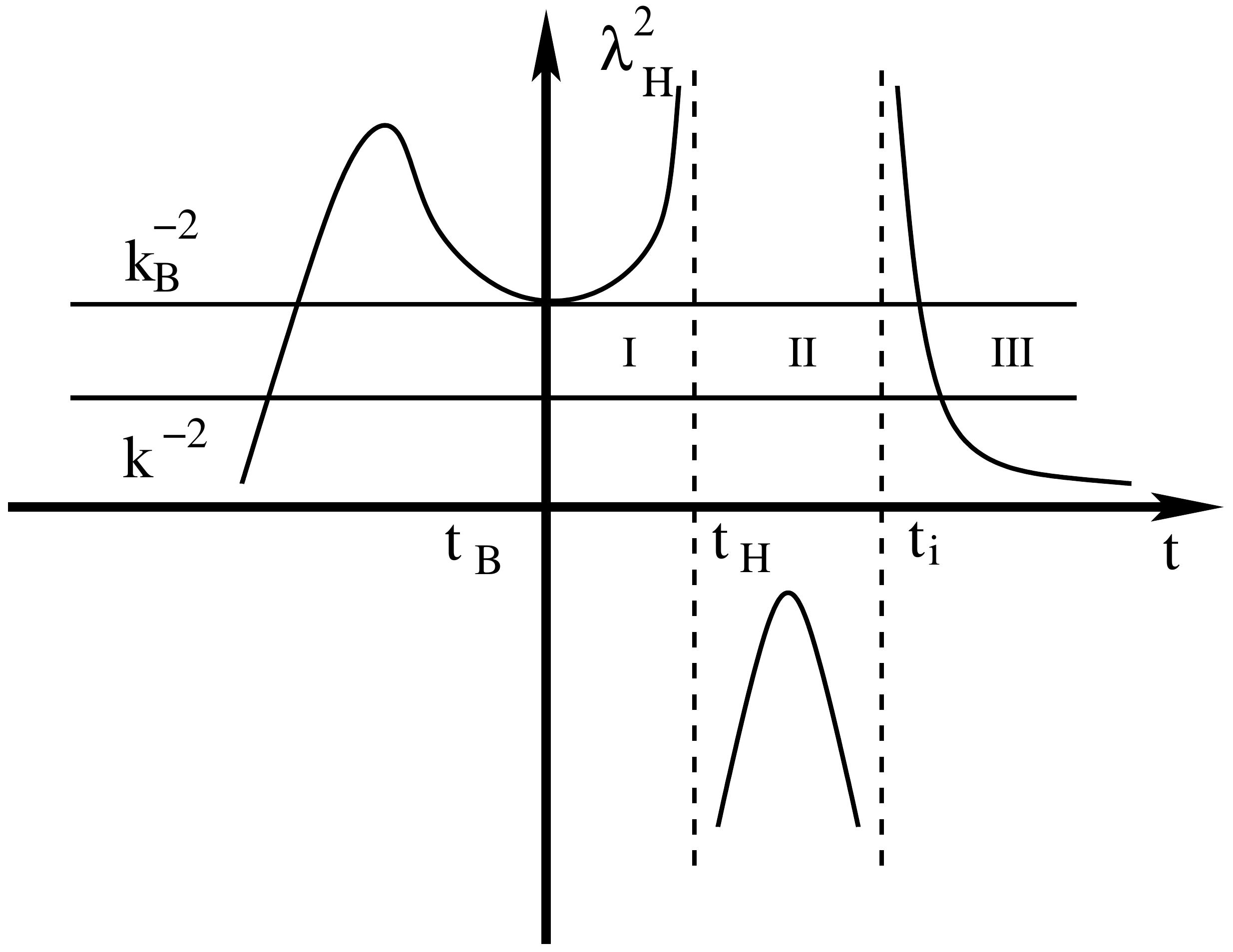} 
}
\caption{Schematic plot of  $\lambda_H^2$ defined by Eq.(\ref{5.8})   vs $t$ for mLQC-I in the dressed metric approach for the quadratic and the Starobinsky potentials,  where
 { $a''(t_{H}) =0$  and $ a''(t_i)=0$ with $t_i$ being the starting time of the inflationary phase.
During the slow-roll inflation, we have $\lambda_H^2\approx L_\text{H}^2/2$. In the contracting phase $t < t_B$, the universe is initially de Sitter  and  we still have $\lambda_H^2 \approx L_\text{H}^2/2$, but now
it  increases exponentially toward bounce $t \rightarrow t_B$, as the universe in this phase is exponentially contracting.}  {However, several Planck seconds before the bounce, the universe enters a non-de Sitter state, during which $\lambda_H^2$ starts to decrease until the bounce}. The qualitative behavior of the comoving Hubble radius is the same for the quadratic and the Starobinsky potentials. Different potentials will change the the values of $t_H$ and $t_i$ correspondingly.}
\label{fig5}
\end{figure} 

It should be noted that if the initial conditions are imposed at the bounce, from Fig. \ref{fig5} we can see clearly that some modes are inside the comoving Hubble radius, and some are not.
 In addition, in the neighborhood of the bounce, the background is far from de Sitter. So,  it is impossible to impose either the BD vacuum or the de Sitter state at the bounce. In this case,   one of the choices of the initial conditions is  the fourth-order adiabatic  vacuum,  similar to that in LQC    \cite{aan2013b,d1,d1b,ZhuA,ZhuB}.

 \subsubsection{mLQC-II}

In mLQC-II, the evolution of the effective homogeneous universe is different from that of mLQC-I. In particular, it is symmetric with respect to the bounce and in the initially kinetic energy dominated
case at the bounce the solutions can be well approximated by Eq.(\ref{3.23}) in the bouncing phase \cite{lsw2018b,lsw2019}, similar to that of LQC \cite{LQC}. 

When considering the cosmological perturbations, similar ambiguities in the choices of $\pi_a^{-2}$ and $\pi_a^{-1}$ in Eq.(\ref{5.9}) exist. In particular, for  the choice of Eq.(\ref{5.11d}) now the function $\Theta(b)$ is replaced by
\bq
\lb{5.16}
\Theta(b) = \cos\left(\frac{\lambda b}{2}\right),
\eq
which  behaves also like a step function across the bounce and picks up the right sign  in both contracting and expanding phases, so it smoothly connects $\Omega_{\pm}$ defined by Eq.(\ref{5.11c}). 

On the other hand, $ \Omega^2_\mathrm{eff}$ is obtained from Eq.(\ref{5.9}) by the replacements,
\bqn
\lb{5.17}
\frac{1}{\pi^2_a}&\rightarrow& \frac{4\pi^2 \gamma^2\lambda^2}{9a^4\sin^2\left(\lambda b/2\right){\cal{D}}} , \\
\frac{1}{\pi_a}&\rightarrow& \frac{-2\pi \gamma \lambda \cos\left(\lambda b/2\right)}{3a^2\sin\left(\lambda b/2\right){\cal{D}}^{1/2}},
\eqn
but now with 
\bq
\lb{5.18}
{\cal{D}} \equiv 1+\gamma^2\sin^2\left(\frac{\lambda b}{2}\right).
\eq

Such obtained  $\Omega^2$, $\Omega^2_{\pm}$  and $ \Omega^2_\mathrm{eff}$ are quite similar to those  given  by Figs. \ref{fig3} and \ref{fig4}  in mLQC-I.  In particular, at the bounce, 
we have   
\bqn
\lb{5.19}
&& \Omega^2(t_B) = 1.59\times 10^{-10}, \quad \Omega^2_\mathrm{eff}(t_B) =0.265,\nb\\
&& k_B^{\scriptscriptstyle{\mathrm{II}}} = \left.\left(\frac{a''}{a}\right)^{1/2}\right|_{t=t_B} \approx 6.84, 
\eqn
that is, the curvature term $a''/a$ still dominates the evolution near the bounce.

\begin{figure}[h!]  
{
\includegraphics[width=7cm]{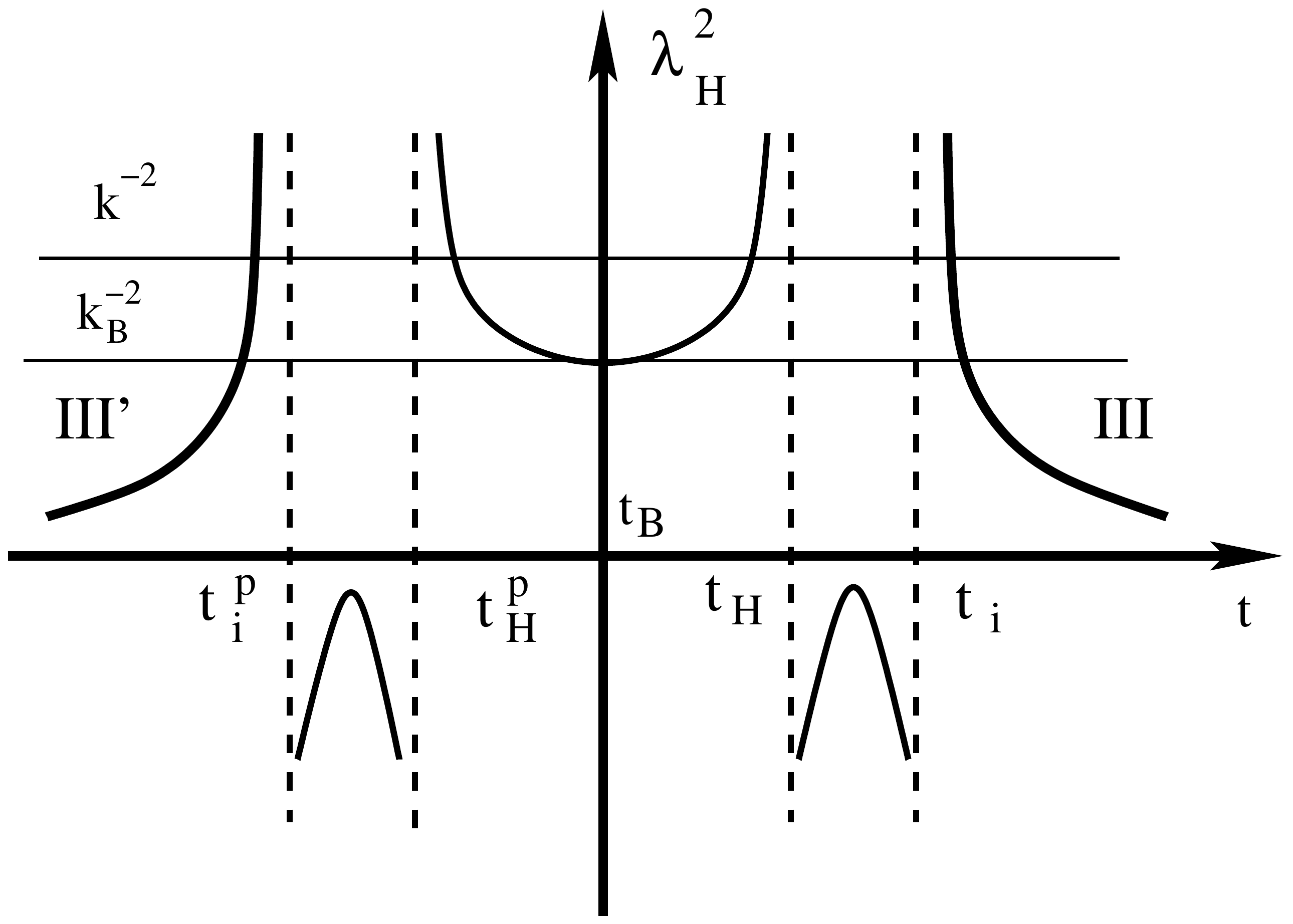}
}
\caption{ {Schematic plot of  $\lambda_H^2$ defined by Eq.(\ref{5.8})   vs $t$ for mLQC-II in the dressed metric approach for the quadratic and the Starobinsky potentials, where $a''(t^p_{H})=a''(t_{H})=0$  and $a''(t^p_i)=a''(t_i)=0$, 
 and $t = t_i$ denotes the starting time of the inflationary phase, while $t = t^p_i$ is the end time of the deflationary phase in the contracting branch.
During the slow-roll inflation, we have $\lambda_H^2\approx L_\text{H}^2/2$. 
In particular,  $\lambda_H^2$ is decreasing (increasing) exponentially in Region III (Region III'). The corresponding effective mass near the bounce is always negative.
Similar behavior also happens in LQC in the dressed metric approach \cite{ZhuA}. The bounce is dominated by the kinetic energy of the scalar field,
which leads to $t^p_H\approx-t_H$. However, in general  we find that $t^p_i \neq  -t_i$ due to the effects of the potential energy of the scalar field far from the bouncing point.} The comoving Hubble radius has the same qualitative behavior for the quadratic and the Starobinsky potentials, while the values of $t_i$ ($t^P_i$) and $t_H$ ($t^p_H$) depend on the type of the potentials and the initial conditions.}
\label{fig6}
\end{figure}

To see how to impose the initial conditions, let us  introduce the quantity $\lambda^2_H$ defined by Eq.(\ref{5.8}) but now $\mathfrak{A}_{-}$ will be replaced either by 
$\Omega^2$ or $ \Omega^2_\mathrm{eff}$.  The details here are not important, and   $\lambda^2_H$ is schematically plotted in Fig. \ref{fig6}, from which we can see that 
if the initial conditions are chosen to be imposed at the bounce,  the BD vacuum (as well as the de Sitter state)  is still not available, and the fourth-order adiabatic vacuum is one of the possible choices, similar to the 
LQC case. However,  if the initial conditions are imposed in the contracting phase at $t_0 \ll t^{p}_i$, the universe becomes very large $a(t) \gg 1$ and can be
practically considered as  flat, then   {the BD vacuum}  can be chosen.

Certainly, one can choose different initial conditions. In particular,  the fourth-order adiabatic vacuum was chosen even in the contracting phase in \cite{lsw2020}. With such a choice, 
  the power spectra from  $\Omega^2$ and  $\Omega^2_\mathrm{eff}$ in the region $k\in(5\times 10^{-6},50)$ were studied and found that the relative difference in the magnitude of the power spectra  is around $30\%$  in the IR regime and less than  $10\%$  in the intermediate regime.  In the UV regime, the relative difference can be as small as $0.1\%$ or even less.

  \subsubsection{LQC}
  
  To consider the effects of the ambiguities in the choice of $\pi_a^{-2}$ and $\pi_a^{-1}$ in Eq.(\ref{5.9}) \footnote{In the framework of LQC, such effects were also studied in
  \cite{aan2013b,ZhuA,ZhuB,nbm2018,lsw2020}.   {In particular, in \cite{abs2018,IA19,AKS20} the function  $\mathfrak{A}_{-}$ defined in Eq.(\ref{5.11a}) was chosen over the whole process 
  of the evolution of the universe.}}, power spectra of the cosmological perturbations were also studied  in the framework of LQC
  in \cite{lsw2020}.
  In this case, $\Omega^2$ is obtained from Eq.(\ref{5.11d}) with
  \bq
  \lb{5.20}
  \Theta(b) = \cos(\lambda b),
  \eq
  while $\Omega^2_\mathrm{eff}$ is obtained from Eq.(\ref{5.9}) by the replacements,
  \bqn
\lb{5.21a}
\frac{1}{\pi^2_a}&\rightarrow& \frac{16\pi^2 G^2 \gamma^2\lambda^2}{9a^4\sin^2\left(\lambda b\right)}, \\
\lb{5.21b}
\frac{1}{\pi_a}&\rightarrow& \frac{-4\pi \gamma \lambda \cos\left(\lambda b\right)}{3a^2\sin\left(\lambda b\right)}.
\eqn

As shown explicitly, the term  {$\Omega^2_{+}$} is always negligible comparing with the curvature term $a''/a$ in the expression of $\Omega^2_{\text{tot}}$ defined in Eq.(\ref{5.7}) by replacing 
$\mathfrak{A}_{-}$ with $\Omega^2_{+}$. So, from Eq.(\ref{5.8}) we find that
 \bqn
  \lb{5.22}
  \lambda_H^2 = \frac{1}{a''/a - \Omega^2_{+}} \simeq \frac{a}{a''},
  \eqn
during the bouncing phase $t \in (t_B, t_i)$, and $ \lambda_H^2 \simeq {a}/{a''}$ was shown schematically  by Fig. 18 in \cite{ZhuB}, which is quite similar to Fig. \ref{fig6} given above for mLQC-II.

As a result, the initial states of the linear perturbations can be either imposed in the contracting phase at a moment   
 {$t_0 \ll t^p_i$} as  {the BD vacuum}, or at the bounce as the fourth-order adiabatic vacuum \cite{aan2013b}. However,  it was shown analytically that such two conditions lead to the same results \cite{ZhuB}. 

\begin{figure}
\includegraphics[width=8cm]{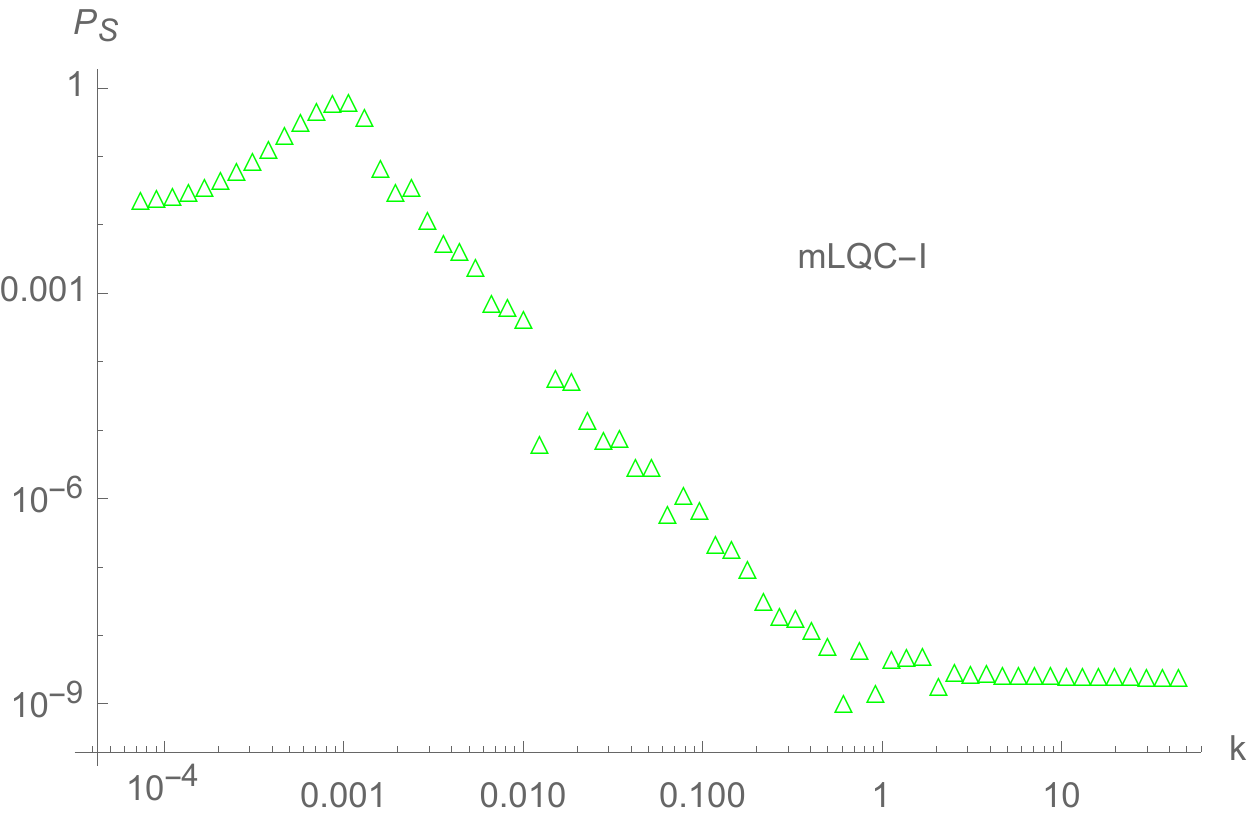}
\includegraphics[width=8cm]{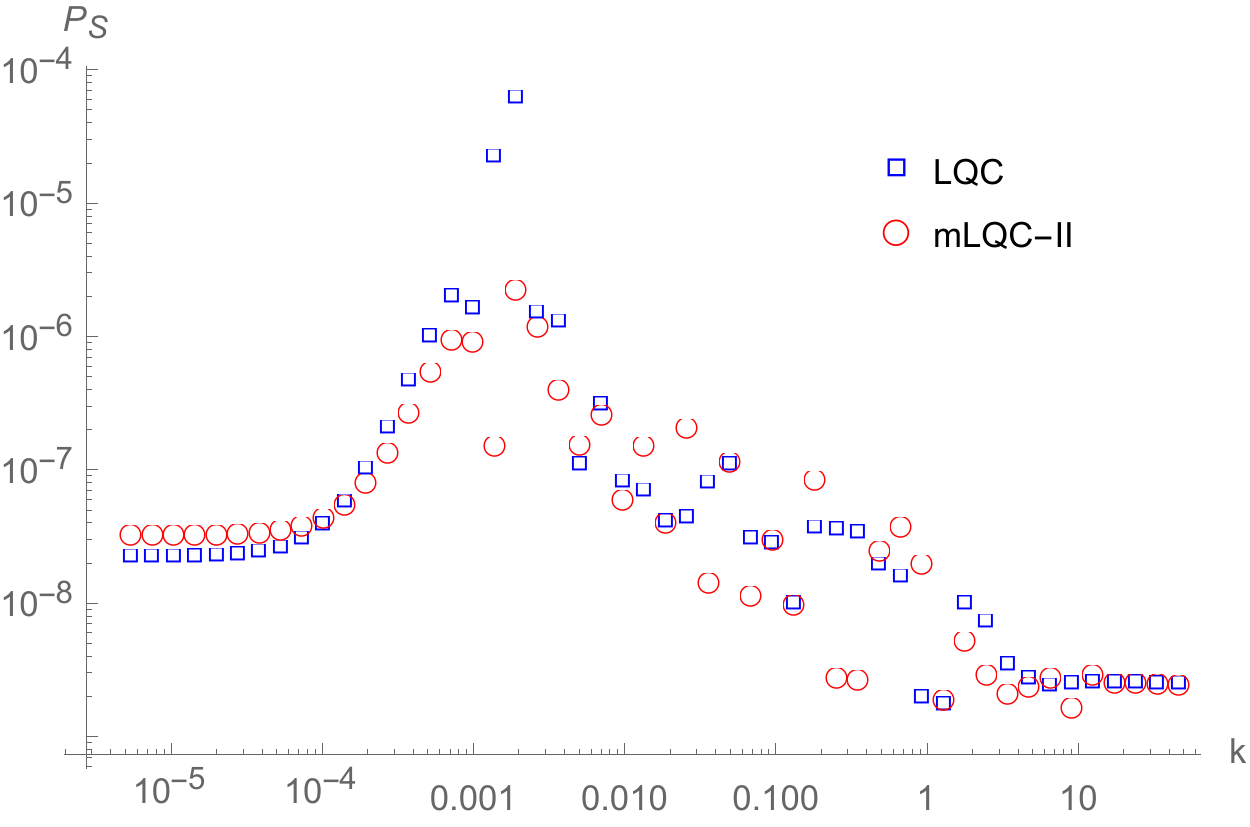}
\caption{The figure shows the results of the scalar power spectra from three models presented in \cite{lsw2020} when the potential term is given by $\Omega_{\text{eff}}^2$. The inflationary potential is chosen to be the  quadratic potential and the e-foldings of the inflationary phases in all three models are chosen to be $72.8$. The first panel shows the scalar power spectrum in mLQC-I which is characterized by its unique infrared regime. In the second panel, we compare the scalar power spectra from LQC and mLQC-II.  }
\label{fig7}
\end{figure}

To compare the results obtained from the three different models, in \cite{lsw2020} the fourth-order adiabatic vacuum was chosen even in the contracting phase for LQC. Here, we cite some of the results in Fig. \ref{fig7}. In particular, it was found that 
the relative difference in the amplitudes of the power spectra of the scalar perturbations due to the choice of $\Omega^2$ or $\Omega^2_\mathrm{eff}$ is about $10\%$ in the infrared regime, about $100\%$
 in the intermediate  regime, and about $0.1\%$ in the UV regime. Since only  modes in the UV regime can be observed currently, clearly this difference is out of the sensitivities of the current and forhcoming 
 observations   \cite{S4-CMB}.  
 
 However, comparing the power spectra obtained from the three different models,  {even with the same choice of $\pi_a^{-2}$ and $\pi_a^{-1}$,}
  it was found that   the relative difference among LQC, mLQC-I and mQLC-II are significant only in
 the  IR and oscillating regimes, while in the UV regime, all three models give quite similar results. In particular, with the same regularization of $\pi_a$ the difference
 can be as large as $100\%$ throughout  the IR and oscillating regimes, while in the UV regime it is about  $0.1\%$. 
 
 For the tensor perturbations, the potential term $\Omega_Q$ vanishes identically, so no ambiguities related to the choice of $\pi_a$ exist. But, due to different models, the differences of the power spectra of the tensor perturbations 
 can be still very large in the IR and oscillating regimes among the three models, although they are very small in the UV regime, see, for example, Fig. 12 given in \cite{lsw2020}.

 \section{Primordial power spectra  of Modified LQCs in  Hybrid Approach}  
\label{SecIV}
\renewcommand{\theequation}{4.\arabic{equation}}\setcounter{equation}{0}

 As in the previous section, in this section we also consider the three different models, LQC, mLQC-I, and mLQC-II, but now in the hybrid approach, and pay particular attention to the 
 differences of the power spectra among these models. Since the scalar perturbations are the most relevant ones in the current CMB observations, in the following we shall mainly focus on
 them, and such studies can be easily extended to the tensor perturbations.

 \subsection {mLQC-I}

Power spectra of the cosmological scalar and tensor perturbations for the effective Hamilton in mLQC-I were recently studied in the hybrid approach  \cite{gqm2020,qmp2020,LOSW20}.
In particular, the   mode function $v_k$ of the scalar perturbations satisfies the differential equation \cite{LOSW20},
 \bq
 \lb{5.23}
v_k''  + \left(k^2 + s\right)v_{ k} = 0,
 \eq
 where 
 \bqn
 \lb{5.24}
s&=&\frac{4 \pi G p^2_\phi}{3 v^{4/3}}\left(19-24 \pi G \gamma^2 \frac{p^2_\phi}{\pi_a^2}\right)\nb\\
&& + v^{2/3}\left(V_{, \phi\phi}+\frac{16 \pi G \gamma p_\phi }{\pi_a}V_{,\phi}-\frac{16\pi G}{3}V\right)\nb\\
&=& - \frac{4\pi G}{3} a^2\left(\rho- 3P\right) + {\cal{U}}, ~~~~
 \eqn
which is the effective mass of the scalar mode, with
\bqn 
\lb{5.24a}
{\cal{U}} &\equiv& a^2\left[V_{,\phi\phi}  - 12 \frac{V_{,\phi}}{\pi_a}\right. \nb\\
&&\left.  + \frac{64 a^6 V(\phi)}{\pi G}\left(\rho - \frac{3V(\phi)}{4\pi G}\right)\frac{1}{\pi_a^2}\right].
\eqn
Note that in  \cite{LOSW20}, instead of $\pi_a$, the symbol  $\Omega$ was used. In addition, the cosmological tensor perturbations are also given by Eqs.(\ref{5.23}) and (\ref{5.24})
but with the vanishing potential  ${\cal{U}} = 0$. Then, we immediately realize that  { in the hybrid approach 
quantum mechanically there are also} ambiguities in the replacements $\pi_a^{-2}$ and  $\pi_a^{-1}$,  as  mentioned in the last section.  So far, two possibilities  were
considered \cite{gqm2020,qmp2020}. One is given by the replacements,  
\bq
\lb{5.25}
\frac{1}{\pi_a^2}\rightarrow \frac{1}{\Omega^2_{{\scriptscriptstyle{\mathrm{I}}}}}, \quad \quad  \frac{1}{\pi_a}\rightarrow \frac{\Lambda_{{\scriptscriptstyle{\mathrm{I}}}}}{\Omega^2_{\scriptscriptstyle{\mathrm{I}}}},
\eq
in Eq.(\ref{5.24}), where 
\bqn
\lb{5.26}
\Omega^2_{{\scriptscriptstyle{\mathrm{I}}}}&\equiv&-\frac{v^2\gamma^2}{\lambda^2}\Big\{\sin^2\left(\lambda b\right)-\frac{\gamma^2+1}{4\gamma^2} \sin^2\left(2 \lambda b\right) \Big\},\nb\\
\Lambda_{{\scriptscriptstyle{\mathrm{I}}}} &\equiv& v\frac{\sin(2\lambda b )}{2\lambda}.
\eqn
This is the case referred to as prescription A in \cite{qmp2020}. 

The other possibility is obtained by the replacement of Eqs.(\ref{5.21a}) and (\ref{5.21b}), which was  referred to as Prescription B  \cite{qmp2020}, 
and showed that the two prescriptions  lead to almost the same results. 
So, in the rest of this  {section} we restrict ourselves only to  prescription A. 

Then,  for the case in which the evolution of the homogeneous universe was dominated by kinetic energy at the bounce,
\bq
\lb{5.27}
\dot\phi^2_B \gg 2V(\phi_B),
\eq
it was shown that the effective mass  is always positive at the bounce  \cite{qmp2020}. In fact, near the bounce we have \cite{WZW18},
\bqn
\lb{5.28}
s &=& - \frac{4\pi G}{3}a^2\left(\rho - 3P\right) + {\cal{U}}(\eta)\nb\\
 &\simeq&  \frac{8\pi G}{3}a^2\rho > 0.
\eqn
Note that in writing the above expression, we have used the fact that during the bouncing phase we have $w_{\phi} \equiv P/\rho \simeq 1$, and $\left|{\cal{U}}(\eta)\right| \ll 1$.  
On the other hand, in the pre-bounce phase,  when $t \ll t_B$ the background is a contracting de Sitter spacetime, so   we have \cite{qmp2020}, 
\bqn
\lb{5.29}
s &=& - \frac{4\pi G}{3}a^2\left(\rho - 3P\right) + {\cal{U}}(\eta)
 \simeq  {\cal{U}}(\eta) \simeq 5 a^2 V_{,\phi\phi},\nb\\
 a &\simeq& a_B e^{H_{\Lambda}(t- t_B)},
\eqn
 where $H_{\Lambda} \equiv - \sqrt{8\pi \alpha G\rho_{\Lambda}/3}$. Thus,   the effective mass remains positive in the pre-bounce phase, as long as 
 $V_{,\phi\phi}(t \ll t_B) > 0$.  This is the case for  {both quadratic and  Starobinsky potentials}. In fact, from (\ref{5.29a}), we find that
   \bq
 \lb{5.29b}
  V_{,\phi\phi}  =\begin{cases}
  m^2 , & \text{quadratic},\cr
  m^2 \left(2-e^{4\sqrt{\pi G/3}\;\phi}\right)e^{-8\sqrt{\pi G/3}\;\phi}, & \text{Starobinsky}.\cr
  \end{cases}
  \eq
  {For the case that satisfies the condition (\ref{5.27}) initially  at the bounce, we find that $\phi(t)$ becomes very negative at $ t \ll t_B$ for the  Starobinsky potential, so $ V_{,\phi\phi}(t \ll t_B)$ is positive even 
 in this case. Then,}  the quantity defined by
 \bq
 \lb{5.30}
 \lambda_H^2 \equiv - \frac{1}{s}, 
 \eq
 has similar behavior in the post-bounce phases for  the case in which the evolution of the homogeneous universe was dominated by kinetic energy at the bounce,
 but has  different behaviors in the pre-bounce phases, depending specifically on the potentials considered. 
 
  In Figs. \ref{fig8} and \ref{fig9} we show the comoving Hubble radius for the  quadratic and Starobinsky potentials, respectively. 
 From these figures it is clear that for $t_i^p < t < t_i$, $\lambda_H^2$ is strictly negative, which implies the effective mass $s$ is positive in this regime. Hence, all the modes assume the oscillatory behavior as the modes inside the Hubble horizon,   and we may impose  the BD vacuum  at the bounce. In addition, when $ t \ll t^p_i$, the background is  well described by the de Sitter space,  so  the de Sitter state   can be imposed in the deep contracting phase. However, imposing  the BD vacuum  at the bounce will clearly lead to different power spectra at the end of the slow-roll inflation from that obtained by imposing
 the  de Sitter state in the deep contracting phase. This is because, when the background is
 contracting to about the moments $t \simeq t^{p-}_i$,  the effective mass becomes so large and negative  that the mode function $v_k$ will be modified significantly, in comparison with that given at $t_0 \;(\ll t^p_i$), or in other words,  particle creation now becomes not negligible during the contracting phase.  Then,  other initial conditions at the bounce may need to be considered.

\begin{figure}[h!]  
{
\includegraphics[width=7cm]{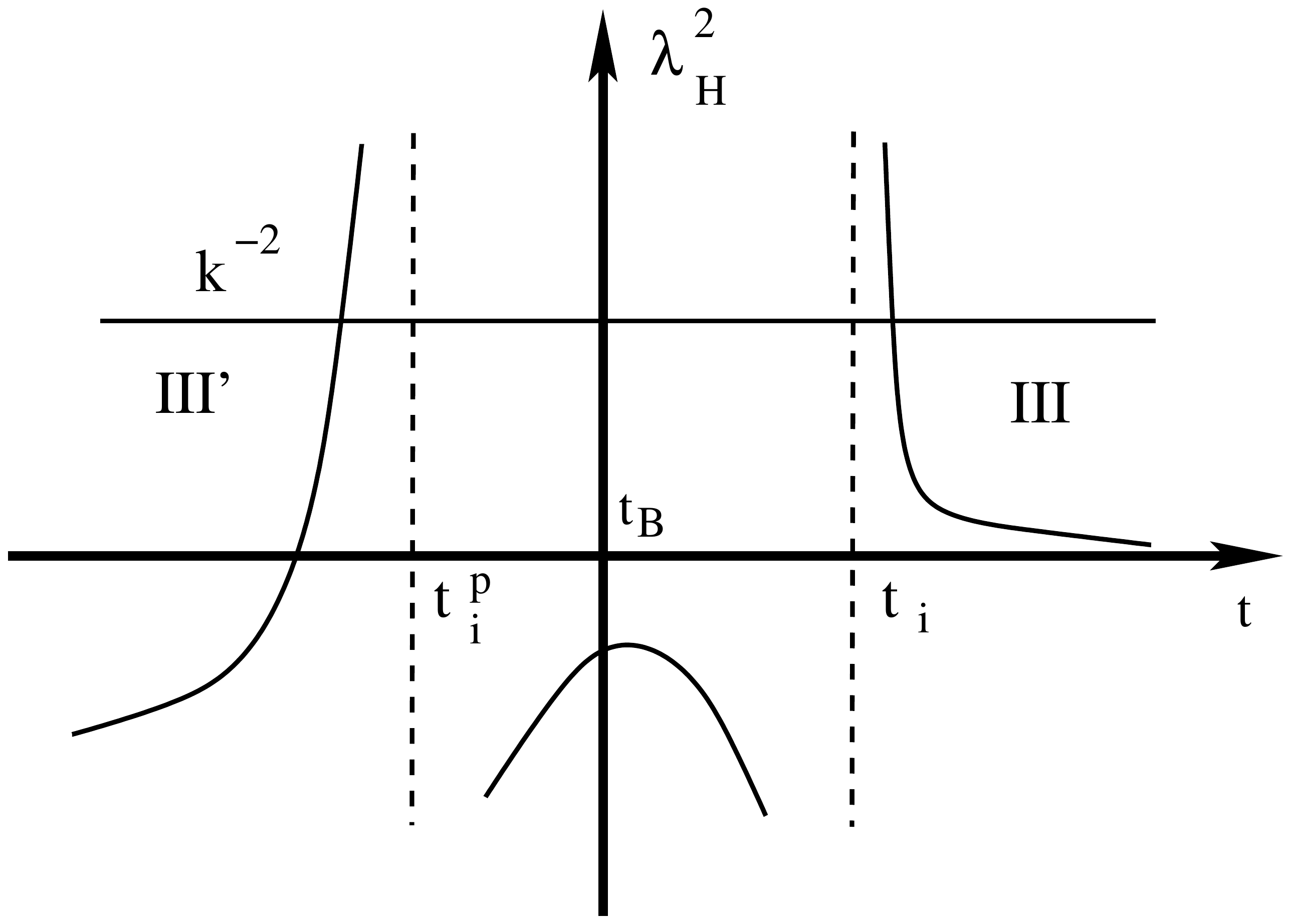}
}
\caption{ {Schematic plot of  $\lambda_H^2$ defined by Eq.(\ref{5.30}) in mLQC-I for the quadratic potential in the hybrid approach, where $s(t_{i}) = s(t^p_i) = 0$, and
 $t = t_i$ is the starting time of the inflationary phase.
During the slow-roll inflation, we have $\lambda_H^2\approx L_\text{H}^2/2$ (Region III).  In the contracting phase, the background is  {asymptotically de Sitter.}
The evolution of the universe is asymmetric with respect to  the bounce. In particular, 
 $\lambda_H^2$    is strictly negative    for $t^p_i < t < t_i$, while for $t \simeq t^{p -}_i$ 
 the ``generalized" comoving Hubble radius $\lambda_H^2$    becomes positive   and large.
 However, as $t$ decreases, $\lambda_H^2$    becomes   negative   again.} Although the values of $t_i$ and $t_H$ depend on the initial conditions for the background evolution, for example, when $\phi_B=1.27~m_\mathrm{pl}$ at the bounce, $t_i\approx7.55\times10^4 ~t_\mathrm{pl}$ and $t_H\approx-21.85~t_\mathrm{pl}$, the qualitative behavior of the comoving Hubble radius is robust with respect to the choice of the initial conditions as long as the bounce is dominated by  the kinetic energy of the scalar field. }
\label{fig8}
\end{figure}

\begin{figure}[h!]  
{
\includegraphics[width=7cm]{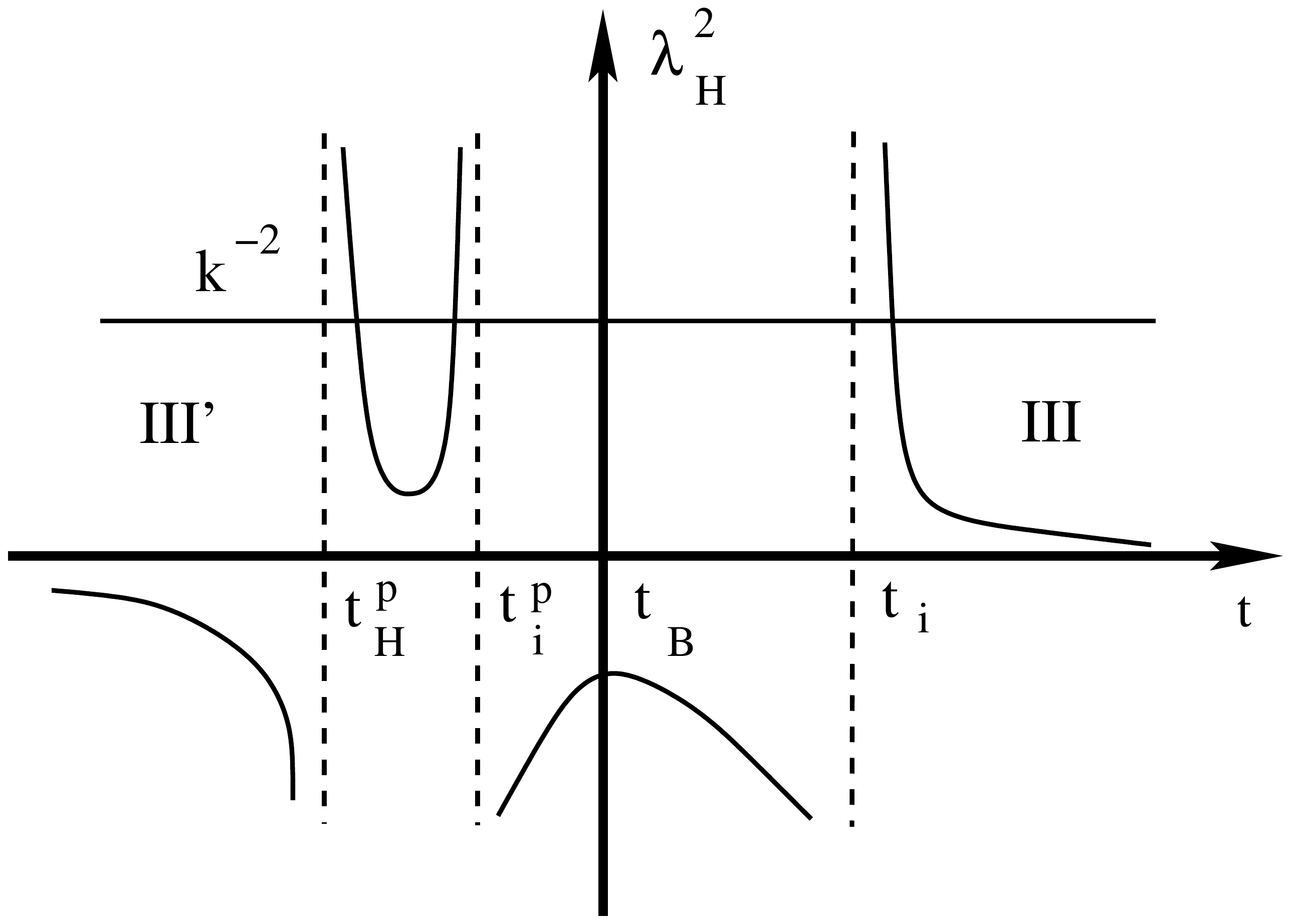}
}
\caption{ {Schematic plot of  $\lambda_H^2$ defined by Eq.(\ref{5.30})   for the Starobinsky potential  and mLQC-I in the hybrid approach, where $s(t_{i}) = s(t^p_i) = s(t^p_H) = 0$, and
 $t = t_i$ is the starting time of the inflationary phase.
During the slow-roll inflation, we have $\lambda_H^2\approx L_\text{H}^2/2$ (Region III).  In the contracting phase, the background is  {asymptotically de Sitter.}
The evolution of the universe is asymmetric with respect to  the bounce. In particular, 
 $\lambda_H^2$  is strictly negative  for $t^p_i < t < t_i$, while for $t \simeq t^{p -}_i$ it   becomes positive  and large.
 However, as $t$ decreases, $\lambda_H^2$   becomes   negative again.} The qualitative behavior of $\lambda^2_H$ does not change with the choice of the initial conditions as long as the inflaton initially starts from the left wing of the potential at the kinetic-energy-dominated bounce. However, the exact values of $t_i$ and $t_H$ depend on the initial conditions. For example, when $\phi_B=-1.32~m_\mathrm{pl}$, $t^p_H=-7.88~t_\mathrm{pl}$, $t^p_i=-4.11~t_\mathrm{pl}$ and $t_i=4.90\times10^5~t_\mathrm{pl}$.}
\label{fig9}
\end{figure}

 \subsection{mLQC-II}
 
 Similar to  LQC, the homogeneous universe of  mLQC-II is symmetric with respect to the bounce, and is well described by the analytical solutions given by Eqs.(\ref{3.23}) and (\ref{3.24})
 for the states that are dominated by kinetic energy at the bounce. 
 
 In this model, the cosmological perturbations are also given by Eqs.(\ref{5.23})-(\ref{5.24a}) but now with the replacement  \cite{LOSW20},
 \bq
\lb{5.31}
\frac{1}{\pi_a^2}\rightarrow \frac{1}{\Omega^2_{{\scriptscriptstyle{\mathrm{II}}}}}, \quad \quad  \frac{1}{\pi_a}\rightarrow \frac{\Lambda_{{\scriptscriptstyle{\mathrm{II}}}}}{\Omega^2_{\scriptscriptstyle{\mathrm{II}}}},
\eq
 where
 \bqn
\lb{5.32}
\Omega^2_{{\scriptscriptstyle{\mathrm{II}}}}&\equiv& \frac{4v^2}{\lambda^2}\sin^2\left(\frac{\lambda b}{2}\right)\left\{1+\gamma^2 \sin^2\left(\frac{\lambda b}{2}\right)\right\},\nb\\
\Lambda_{{\scriptscriptstyle{\mathrm{II}}}} &\equiv& v \frac{\sin\left(\lambda b\right)}{\lambda}.
\eqn

In this case, it can be shown that the effective mass defined by Eq.(\ref{5.24}) is always positive in the neighborhood of the bounce, but  far from the bounce,  the properties of $\lambda^2_H$
depend on the potential in the pre-bounce phase, similar to  mLQC-I. 

 In Fig. \ref{fig10}, we plot $\lambda^2_H$ for the  Starobinsky potential, while for the quadratic one, it is quite similar to the corresponding one in mLQC-I, given by
Fig. \ref{fig8}.  From Fig. \ref{fig10}  we can see that $\lambda^2_H$ now is negative not only near the bounce but also in the whole contracting phase,
 so that   all the modes are oscillating for $t < t_i$. Then, one can choose  the BD vacuum at the bounce. It is remarkable that for the quadratic potential, this is impossible [cf. Fig. \ref{fig8}].
 
 Moreover, as $t \rightarrow - \infty$, the expansion factor becomes very large, and the corresponding curvature is quite  low, so to a good approximation, 
 the BD vacuum  can also be chosen in the distant past, not only for the Starobinsky potential but also for other potentials. Due to the oscillating behavior of the mode function over the whole contracting phase, 
imposing  the BD vacuum  at the bounce is expected not to lead to  significant difference in the power spectra from that in which the same condition is imposed in the deep contracting phase.

\begin{figure}[h!]  
{
\includegraphics[width=7cm]{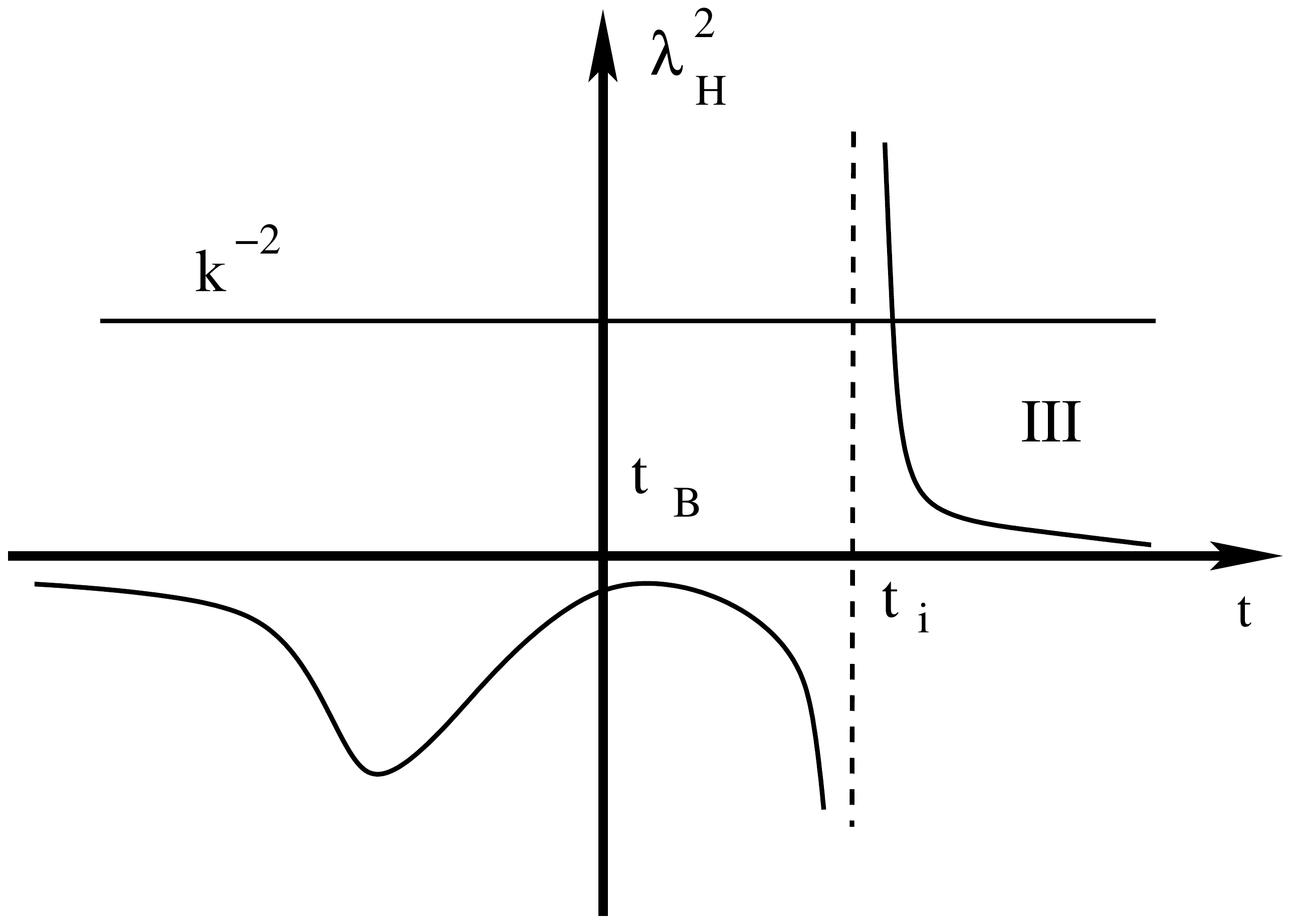}
}

\caption{ Schematic plot of  $\lambda_H^2$ defined by Eq.(\ref{5.30})   for the Starobinsky  potential  and mLQC-II in the hybrid approach, where $s(t_{i}) = 0$, and
 $t = t_i$ is the starting time of the inflationary phase.
During the slow-roll inflation, we have $\lambda_H^2\approx L_\text{H}^2/2$ (Region III) since the contribution from the potential is in general less than $a''/a$. 
Again the qualitative behavior of $\lambda^2_H$ remains the same as long as the inflaton starts from the left wing of the potential  with a positive velocity and
 the bounce is initially dominated by the kinetic energy of the inflaton field.}
\label{fig10}
\end{figure}

 \subsection{LQC}
 
 The evolution of the homogeneous universe of  standard  LQC model is also symmetric with respect to the bounce, and is well described by the analytical solutions given in \cite{ZhuA,ZhuB}
 for the states that are dominated by kinetic energy at the bounce. 
 
 In this model, the cosmological perturbations are also given by Eqs.(\ref{5.23})-(\ref{5.24a}) but now with the replacement  \cite{LOSW20},
 \bq
\lb{5.33a}
\frac{1}{\pi_a^2}\rightarrow \frac{1}{\Omega^2_{\text{LQC}}}, \quad \quad  \frac{1}{\pi_a}\rightarrow \frac{\Lambda_{\text{LQC}}}{\Omega^2_{LQC}},
\eq
 where
 \bqn
\lb{5.34a}
\Omega_{\text{LQC}}&\equiv& \frac{v \sin\left(\lambda b\right)}{\lambda},\;\;\;
\Lambda_{\text{LQC}} \equiv \frac{v \sin\left(2\lambda b\right)}{2\lambda}.
\eqn

 In this case, it can be shown that the effective mass defined by Eq.(\ref{5.24}) is always positive for the states  that are dominated by kinetic energy at the bounce \cite{nbm2018,WZW18}, and   the quantity $\lambda^2_H$ defined by Eq.(\ref{5.30}) is negative near the bounce.  Again, similar to the mLQC-II case,  the modes are oscillating near the bounce.  However,  in the contracting phase the
 behavior of  $\lambda^2_H$ sensitively depends on the inflation potentials. For the Starobinsky one, $\lambda^2_H$ behaves similar to that described by Fig. \ref{fig10}, so the BD vacuum  can be imposed either in the deep contracting phase or at the bounce, and such resulted power spectra are expected not to be significantly different from one another. But for the quadratic potential the situation is quite different, and a preferred choice is to impose  the BD vacuum in the deep contracting phase ($t_0 \ll t_B$).

 \subsection {Primordial Power Spectra}

 As it can be seen that one of the preferred moments to impose the initial conditions for the cosmological perturbations in all these three models is a moment in the contracting phase $t_0 < t_B$. In this phase, we can impose the BD vacuum state  as long as the moment is sufficiently earlier, $t_0 \ll t_B$. Certainly, other initial conditions can also be chosen. In particular, in \cite{LOSW20} the second-order adiabatic vacuum conditions were selected, but  it was found that the same results can also be obtained even when the BD vacuum  or the fourth-order adiabatic vacuum is imposed initially.

   The  nth-order adiabatic vacuum conditions can be obtained as follows:  Let us first consider the solution, 
 \bq
\lb{5.33}
\nu_k=\frac{1}{\sqrt{2 W_k}}e^{-i \int^\eta W_k(\bar \eta)d\bar \eta}.
\eq
Then, inserting it  into (\ref{5.23}), one can find an iterative equation for $W_k$. In particular, it can be shown that  the zeroth-order  solution is given by $W^{(0)}_k=k$, while
 the  second and fourth order adiabatic solutions are given by, 
\bq
\lb{5.34}
W^{(2)}_k=\sqrt{k^2+s}, \quad \quad  W^{(4)}_k=\frac{\sqrt{f(s,k)}}{4|k^2+s|}.
\eq
Here $f(s,k)=5s'^2+16k^4(k^2+3s)+16s^2(3k^2+s)-4s^{''}(s+k^2)$.
 It should be noted that, in order to compare directly with observations,
 it is found convenient to calculate  the power spectrum of the comoving curvature perturbation ${\cal R}_k$, which is related to the Mukhanov-Sasaki variable via
 the relation  ${\cal R}_k=\nu_k/z$, with $z=a\dot \phi/H$. Then,  its power spectrum reads
\bq
\lb{5.37}
\mathcal P_{{\cal R}_k}=\frac{\mathcal P_{\nu_k}}{z^2}=\frac{k^3}{2\pi^2}\frac{|\nu_k|^2}{z^2}.
\eq
In addition, the power spectrum is normally evaluated at the end of inflation, at which all the relevant modes are well outside the Hubble horizon [cf. Fig. \ref{fig2}].

 It should be also noted that  the above formula is only applicable to the case where  $W^{(2)}_k$ and/or $W^{(4)}_k$ remains  real at the initial time. This is equivalent to require
$k^2+s\ge 0$ for $W^{(2)}_k$ and $f(s,k)\ge 0$  for $W^{(4)}_k$. Since the effective mass $s$ in general depends on $t$, it is clear 
that the validity of (\ref{5.34})  depends not only on the initial states but also on the initial times.

 \begin{figure}
\includegraphics[width=8cm]{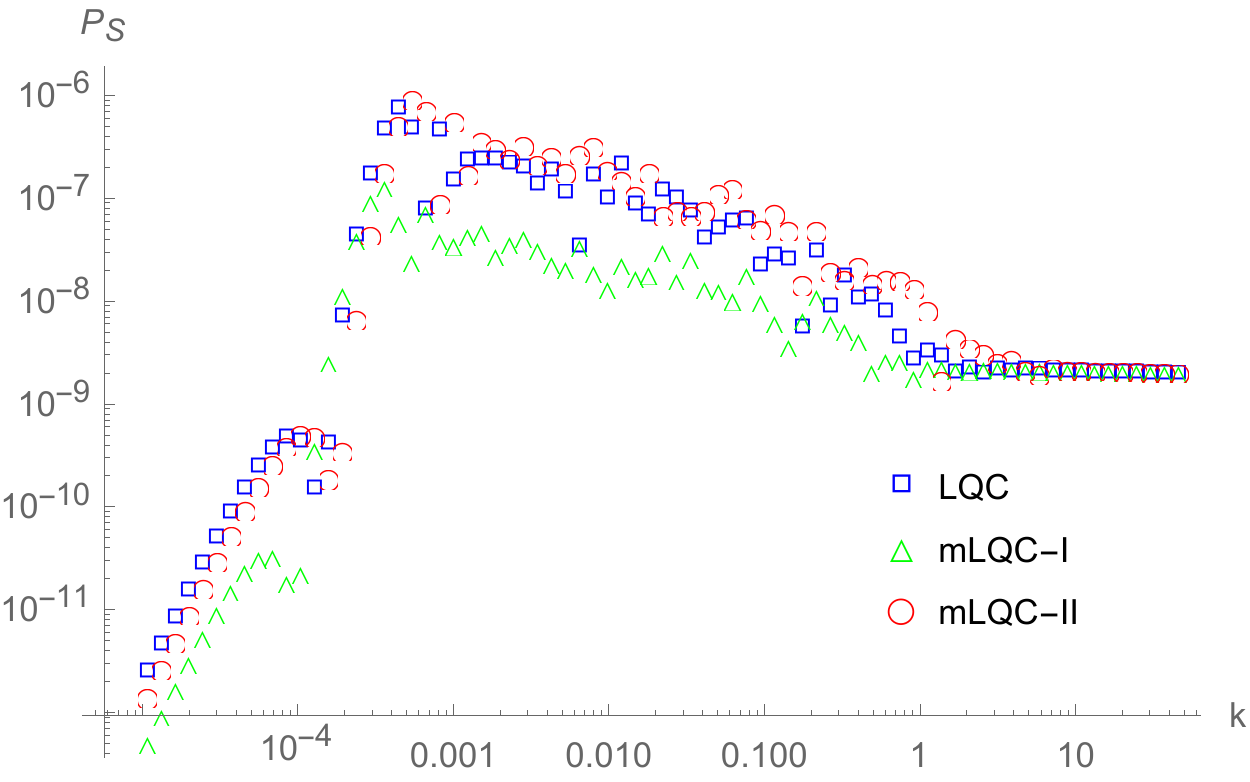}
\caption{The primordial   power spectra of the cosmological scalar perturbations  in the hybrid approach with the Starobinsky potential, respectively, for  LQC, mLQC-I, and mLQC-II. The mass of the inflaton field is set to $2.44\times 10^{-6} m_\mathrm{pl}$. The background evolution is chosen so that the pivot mode is $k_*=5.15$ in all three models. The initial states are the second-order adiabatic states
 imposed in the contracting phase at the moment $t_0$ with  $t_0 \ll t_B$ \cite{LOSW20}.}
\label{fig11}
\end{figure}

In addition,  in the following only the  Starobinsky potential given in Eq.(\ref{5.29a}) will be considered, as it represents one of the most favorable models by current observations \cite{cosmo}. Let us turn to consider the power spectrum of the scalar perturbations in each of the three models. Similar results can be also obtained for the tensor perturbations. In particular, 
it was found that the scalar power spectra in these three models can be still divided into three   distinctive regimes: the  infrared, oscillatory and UV, as shown in Fig. \ref{fig11}.

In the infrared and  oscillatory regimes, the relative difference between LQC and mLQC-I  can be as  large as $100\%$,  while this difference reduces  to less than $1\%$ in the UV regime. This is 
mainly because LQC and mLQC-I  have the same classical limit in the post-bounce phase,  and as shown in Figs. \ref{fig5} and  \ref{fig8}, the effective masses in both approaches  tend to  {be} the same during the  inflationary phase.

However,  it is interesting to note that in the infrared and  oscillatory regimes, the power spectrum in mLQC-I is suppressed  in  comparison  with that of LQC, which  has been
 found only  in the hybrid approach. As a matter of fact, in the dressed metric approach, the power spectrum in mLQC-I is largely amplified in the  infrared regime,
 and  its magnitude is  of the Planck scale   as depicted in Fig. \ref{fig7} \cite{IA19,lsw2020}. The main reason might root in the distinctive behavior of the effective masses in the two approaches,
  as shown explicitly in Figs. \ref{fig5} and \ref{fig8}.

On the other hand,   the  difference of the power spectra between LQC and mLQC-II  is smaller than that between LQC and mLQC-I. In particular, in the  infrared  regime, it is about $50\%$. The large relative difference   
 (more than $100\%$)  of the power spectra between mLQC-I and mLQC-II also happens  in the   infrared and oscillatory regimes,  while  in the UV regime it reduces to about $2\%$.

To summarize, in the hybrid approach  the maximum relative differences of the power spectra among these three different models always happen in the infrared and oscillatory regimes,  while in the UV regime, 
the differences  reduce to no larger than $2\%$, and all the three models predict a scale-invariant power spectrum, and is consistent with the current CMB observations. However,  in the hybrid approach, the power spectrum in mLQC-I is suppressed in the infrared and oscillatory regimes. The latter is   in a striking contrast to the results obtained
from the dressed metric approach, which might be closely related to the fact that  the effective masses in these two approaches are significantly different, especially  near 
the bounce  {and in the prebounce stage}.

 \section{Conclusions and outlook}  
\label{SecV}
\renewcommand{\theequation}{5.\arabic{equation}}\setcounter{equation}{0}

In the past two decades, LQC has been studied extensively, and several remarkable features have been found \cite{LQC}, including the generic resolution of the big bang singularity 
(replaced by a quantum bounce) in the Planckian scale,  the slow-roll inflation as an attractor in the post-bounce evolution of the universe, and the scale-invariant power spectra of the cosmological
perturbations, which are consistent  with  the current CMB observations. Even more interestingly, it was shown recently that  some anomalies from the CMB data \cite{Planck2018b,Planck2019,SCHS16} can be
  reconciled  purely due to the quantum geometric effects in the framework of LQC \cite{AGJS20,AKS20}. 

 Despite of all these achievements, LQC is still plagued with some ambiguities in the quantization procedure. In particular,  its connection with LQG is still not established \cite{engle}, and  
  the quantization procedure used in LQC owing to symmetry reduction before quantization can result in  different Hamiltonian constraints than the one of LQG.

 Motivated by the above considerations, recently various modified LQC models have been proposed, see, for example,  \cite{wilson2017,GFT,QRLG,DID,KT,GLS20a,GLS20b,LSS20,HLL20,ST20}
  and references therein. In this brief review, we have restricted ourselves only to mLQC-I and mLQC-II  \cite{YDM09,DL17,DL18}, as they are the ones that have been extensively
  studied in the literature not only the dynamics of the homogeneous universe  \cite{lsw2018,lsw2018b,lsw2019,SS19a,SS19b,qm2019,gm20}, but also the cosmological perturbations   \cite{IA19,lsw2020,LOSW20,gqm2020,qmp2020}.
  
  In these two modified LQC models, it was found that the resolution of the big bang singularity is also generic  \cite{lsw2018,lsw2018b,lsw2019,SS19a,SS19b}. This is closely related to the fact that 
  the area operator in LQG has a minimal but nonzero eigenvalue \cite{LQG,LQC}, quite similar to the eigenvalue of the ground state of the energy operator of a simple harmonic oscillator in quantum 
  mechanics.  This deep connection  also shows that  the resolution of the big bang singularity is purely due to the quantum geometric effects. In addition, similar to LQC, the slow-roll inflation also occurs generically in both mLQC-I and mLQC-II  \cite{lsw2019}.
  In particular, when the inflaton has a quadratic potential, $V(\phi) = m^2\phi^2/2$, the  probabilities for the desired slow-roll inflation not to occur are $\lesssim 
1.12 \times 10^{-5}$, $\lesssim 2.62 \times 10^{-6}$, and $\lesssim 2.74 \times 10^{-6}$ for mLQC-I, mLQC-II and LQC, respectively.

When dealing with perturbations, another ambiguity rises in the replacement of the momentum conjugate $\pi_a$ of the expansion factor $a$ in the effective potential of the scalar perturbations.
This ambiguity occurs not only in the dressed metric approach [cf. Eq.(\ref{5.9})] but also in the hybrid approach [cf. Eq.(\ref{5.24a})], as it is closely related to the quantization strategy used in LQG/LQC, 
because  now only the holonomies (complex exponentials) of $\pi_a$ are defined as operators.  Several choices have been proposed in the literature  \cite{aan2013b,abs2018,MOP11,IA19,lsw2020,LOSW20,gqm2020,qmp2020}.
In Secs. \ref{SecIII} and  \ref{SecIV}, we have shown that for some choices the effects on the power spectra are non-trivial, while for others the effects are negligible. However, even with the same choice,
 the relative differences in the amplitudes of the power spectra among the three different models can be as large as $100\%$ in the infrared and intermediate regimes of the spectra, while in the UV regime
 the relative differences are no larger than $2\%$, and the corresponding power spectra are scale-invariant. Since only the modes in the UV regime are relevant to the current observations, the power spectra 
 obtained in all the three models are  consistent with current observations \cite{cosmo}.  
  
  However, the interactions between the infrared and UV modes appearing in non-Gaussianities might provide an excellent window to observe such effects.
  This was initially done in LQC \cite{abs2018,ZWKCS18,WZW18}, and lately generalized to bouncing cosmologies \cite{AKS20}. It should be noted that in \cite{AKS20}, the expansion factor
  $a(t)$ near the bounce was assumed to take the form, 
  $$
  a(t) = a_B\left(1 + bt^2\right)^n, 
  $$
  where $b$ and $n$ are two free parameters. For example, for LQC we have $n = 1/6$ and $b = R_B/2$,
  where $R_B$ is the Ricci scalar at the bounce \cite{ZhuA,ZhuB}. But, it is clear that near the bounce $a(t)$ takes forms different from the above expression for
  mLQC-I/II, as one can see from Eqs.(\ref{3.21})-(\ref{3.24}). Thus, it would be very interesting to study such effects in mLQC-I/II, 
  and look for some observational signals. 
  
 Moreover,    initial conditions are another subtle and important issue not only in LQC but also in mLQCs.  As a matter of fact, the initial conditions  consist of two parts:  the initial  time, and 
the initial conditions. Different choices of the initial times lead to different choices of the initial conditions, or vice versa. To clarify these issues, in Sections III and IV we have discussed  it at length by showing the  (generalized) comoving Hubble radius in each model
 as well as  in each of  the two  approaches, dressed metric and  hybrid. From these analyses, we have  shown clearly   which initial conditions can and cannot be imposed at a given initial time.

  In addition, when the universe changes from contraction to expansion at the bounce, particle and entropy 
  creations are expected to be very large, and it is crucial to keep such creations under control, so that the basic assumptions of the models are valid, including the one that the cosmological perturbations are small
  and can be treated  as test fields propagating on the quantum homogeneous background, as assumed  in both the dressed metric and hybrid approaches. 
  
  Yet, different initial conditions also affect the amplitudes and shapes of the primordial power spectra, and it would be very interesting  to investigate the consistency of such obtained spectra 
  with current observations, in particular the possible explanations to the  anomalies found in the CMB data \cite{Planck2018b,Planck2019,SCHS16}, and the naturalness of
  such initial conditions. 
  
  On the other hand, bouncing cosmologies, as an alternative to the cosmic inflation paradigm, have been extensively studied in the literature, see, for example, \cite{BP17} and references therein. However,
  in such classical bounces, exotic matter fields are required in order to keep the bounce open. This in turn raises the question of instabilities of the models. On the other hand, quantum bounces found in 
  LQC/mLQCs are purely due to the quantum geometric effects, and the instability problem is automatically out of the question. So, it would be very interesting to study bouncing cosmologies in the framework of
  LQC/mLQCs. The first step in this direction has already been taken \cite{LSS20}, and more detailed and extensive analyses are still needed.


\section*{Acknowledgements}
We are grateful to Robert Brandenberger and David Wands for valuable comments on the manuscript and helpful discussions. We thank Javier Olmedo and Tao Zhu for various 
discussions related to works presented in this manuscript.    B.-F.L and P.S. are supported by NSF grant PHY-1454832.
 B.-F.L is also partially supported  by  the National Natural Science Foundation of China (NNSFC) with the Grants No. 12005186.

 \end{document}